\documentclass[lettersize,journal]{IEEEtran}
\usepackage{amsmath}
\usepackage{amsthm}
\usepackage{amssymb}
\usepackage{algorithmic}
\usepackage{algorithm}
\usepackage{array}
\usepackage[caption=false,font=normalsize,labelfont=sf,textfont=sf]{subfig}
\usepackage{textcomp}
\usepackage{stfloats}
\usepackage{balance}
\usepackage{url}
\usepackage{verbatim}
\usepackage{graphicx}
\usepackage{thmtools}
\usepackage{arydshln}
\usepackage{cite}
\usepackage{textcomp}
\usepackage[numbers,sort&compress]{natbib}
\renewcommand{\baselinestretch}{1}

\begin{document}

\title{Economic Optimal Power Management of Second-Life Battery Energy Storage Systems}

\author{Amir Farakhor,~\IEEEmembership{Graduate Student Member,~IEEE}, Di Wu,~\IEEEmembership{Senior Member,~IEEE},
		Pingen Chen, Junmin Wang,~\IEEEmembership{Fellow,~IEEE}, Yebin Wang,~\IEEEmembership{Senior Member,~IEEE}, and Huazhen Fang,~\IEEEmembership{Member,~IEEE}
\thanks{A. Farakhor and H. Fang (corresponding author) are with the Department of Mechanical Engineering, University of Kansas, Lawrence, KS, USA (Email: fang@ku.edu, a.farakhor@ku.edu).} 
\thanks{D. Wu is with the Pacific Northwest National Laboratory, Richland, WA, USA (Email: di.wu@pnnl.gov).}
\thanks{Pingen Chen is with the Department of Mechanical Engineering at Tennessee Technological University, Cookeville, TN, USA (Email: pchen@tntech.edu).}
\thanks{ Junmin Wang is with the Walker Department of Mechanical Engineering, University of Texas at Austin, Austin, TX, USA (Email: jwang@austin.utexas.edu).}
\thanks{Y. Wang is with the Mitsubishi Electric Research Laboratories, Cambridge, MA, USA (Email: yebinwang@merl.com).}
}
\markboth{}
{Shell \MakeLowercase{\textit{et al.}}: A Sample Article Using IEEEtran.cls for IEEE Journals}

\maketitle

\begin{abstract}
Second-life battery energy storage systems (SL-BESS) are an economical means of long-duration grid energy storage. They utilize retired battery packs from electric vehicles to store and provide electrical energy at the utility scale. However, they pose critical challenges in achieving optimal utilization and extending their remaining useful life. These complications primarily result from the constituent battery packs’ inherent heterogeneities in terms of their size, chemistry, and degradation. This paper proposes an economic optimal power management approach to ensure the cost-minimized operation of SL-BESS while adhering to safety regulations and maintaining a balance between the power supply and demand. The proposed approach takes into account the costs associated with the degradation, energy loss, and decommissioning of the battery packs. In particular, we capture the degradation costs of the retired battery packs through a weighted average Ah-throughput aging model. The presented model allows us to quantify the capacity fading for second-life battery packs for different operating temperatures and C-rates. To evaluate the performance of the proposed approach, we conduct extensive simulations on a SL-BESS consisting of various heterogeneous retired battery packs in the context of grid operation. The results offer novel insights into SL-BESS operation and highlight the importance of prudent power management to ensure economically optimal utilization.
\end{abstract}


\section{Introduction}
\IEEEPARstart{T}{he} rapidly advancing field of electrified transportation has recently experienced an exponential growth in electric vehicles (EVs), with the trend poised to continue for the upcoming decade. EV lithium-ion batteries typically have a lifespan of about ten years before retirement to ensure vehicle performance and safety. The strong uptake of EVs thus will in turn lead to a substantial number of retired EV batteries, which will exceed 200 GWh per year by 2030 \cite{McKinsey}. These batteries, however, still retain around 70-80\% of the original capacity when reaching the end of their first life \cite{Enviro-LCC-2019}, making them viable for a second life \cite{CellReports-JZ-2021, Review-XG-2024}. The utilization of second-life batteries will bring economic values, enhance environmental sustainability, and strengthen battery supply chains, as increasingly recognized \cite{EnergyConversion-AB-2022}. A particularly promising application, among the various potential avenues, is stationary energy storage systems. These systems present lower demands in cycle life and power and play increasingly important roles in enhancing the reliability and efficiency of the power grid as well as renewable energy facilities. 

A few studies have rallied around the use of second-life battery energy storage systems (SL-BESS). In \cite{EVC-KA-2012, ECCE-KC-2015, ICIT-SI-2015, EEEIC-SA-2021}, SL-BESS is considered for applications that extend from peak load shaving for commercial buildings to residential demand response and to integration of photovoltaic (PV) energy. The research therein investigates the issues of optimal sizing, state-of-charge (SoC) estimation, and rule-based control of SL-BESS. The studies in \cite{AppliedEnergy-MG-2021, IET-DY-2022} deal with energy management for microgrids including SL-BESS, both proposing hierarchical optimization strategies to overcome the complexity in power planning and operation. The work in \cite{TSE-DY-2021} pursues the integration and optimal operation of SL-BESS, PV, and grid for centralized charging infrastructure for EVs. The techno-economic analysis in \cite{AppliedEnergy-ZS-2019} examines the economic benefits and challenges of using SL-BESS for wind farms. These studies adopt a system-of-systems perspective, treating SL-BESS as an essential component within an integrated energy system and pursing comprehensive optimization for the entire system. In a departure, some other studies focus on the operation of individual SL-BESS, and a main subject of inquiry is how to overcome the inhomogeneities among the battery packs in SL-BESS. Active SoC balancing strategies are studied in \cite{SelectedTPEL-RM-2023, Energies-AM-2016}, which align the power allocation with the SoC of every pack. The works in \cite{TPEL-LC-2020, TIE-MN-2015} perform joint power electronics design and control for SoC balancing for SL-BESS. In \cite{TSG-ZQ-2024, APEC-LC-2020}, power allocation is tied with the state-of-health (SoH) of the packs to prolong the overall lifetime of the SL-BESS. 

Existing studies have yielded valuable insights into understanding and utilizing SL-BESS, but some notable limitations persist. The primary challenge to effectively using SL-BESS lies in the heterogeneity of constituent battery packs. The packs, sourced from a vast range of EVs, could have substantially different electrochemistry, energy capacity, power capability, aging condition, and remaining useful life. While it is possible to disassemble the packs and then sort and repackage the cells to build homogeneous SL-BESS, the time and labor costs will be unaffordable. Otherwise, if packs are directly reassembled without homogenization and repurposed for large-scale stationary energy storage, their diverse characteristics will defy treating SL-BESS as a lumped component as in \cite{AppliedEnergy-MG-2021, IET-DY-2022, AppliedEnergy-ZS-2019, TSE-DY-2021}. Further, although the studies in \cite{SelectedTPEL-RM-2023, TPEL-LC-2020, Energies-AM-2016, TIE-MN-2015, APEC-LC-2020} recognize the issue of heterogeneity, it remains unclear whether and how SoC or SoH balancing will benefit the operation of SL-BESS. For instance, SoC balancing can hardly help with packs with varying capacities and aging conditions, and SoH balancing will keep old battery packs idling until all packs achieve the same aging level.

In this paper, we propose to perform economic optimal power management for SL-BESS. We put forward the idea that the primary goal for the operation of SL-BESS should be the maximization of the economic benefits. In line with the goal, we develop an approach to minimize the overall operating costs of SL-BESS while harnessing the heterogeneity to tailor the power management to individual packs according to their characteristics. Our proposed approach specifically considers three key costs, which pertain to the packs' degradation, energy loss, and decommissioning. We present mathematical models to describe these costs. Especially, we introduce an Ah-throughput aging model to grasp the battery packs’ degradation costs under different operating temperatures and charging/discharging C-rates. We formulate and solve an economic optimization problem based on the cost models to achieve optimal power management for SL-BESS. Extensive simulations validate the efficacy of the proposed approach in improving the economic performance by large margins. 

The remainder of this paper is organized as follows. Section II provides an overview of the SL-BESS architecture and outlines the economic criteria for its operation. Section III introduces the electrical, thermal, degradation, and cost models for the SL-BESS and then formulates the proposed economic optimal power management problem. Section IV presents the techno-economic simulation results, and the paper concludes in Section V with final remarks.

\begin{figure}[!t]
\centering
\includegraphics[width=\linewidth]{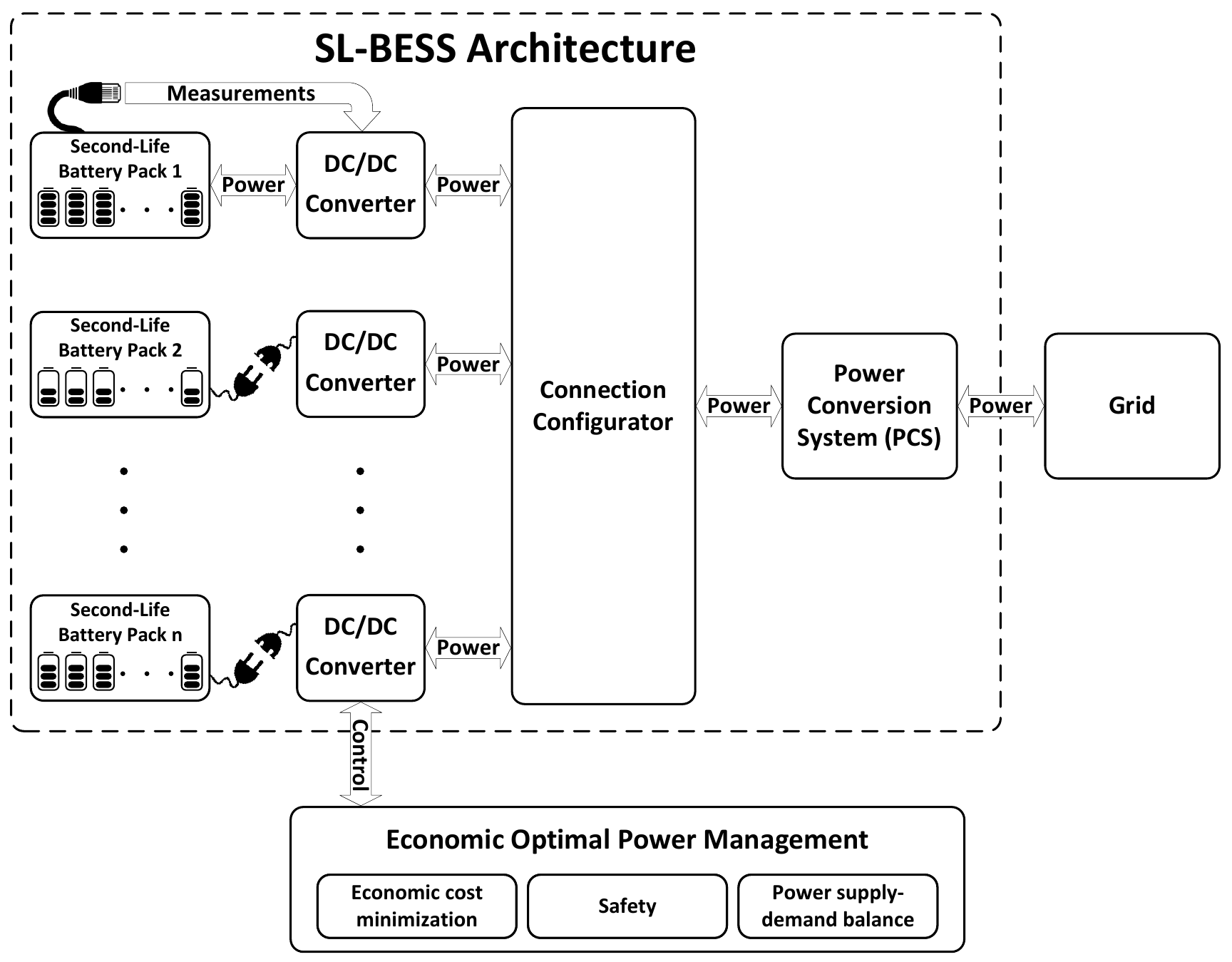}
\caption{The considered SL-BESS design for grid energy storage.}
\label{CircuitStructure}
\end{figure}

\section{Second-Life BESS: An Overview}
This section outlines an SL-BESS paradigm for grid energy storage. We first present the SL-BESS architecture and functionalities and then introduce the economic criteria for its operation. 

\subsection{The SL-BESS Architecture}

We propose the SL-BESS architecture as shown in Fig.~\ref{CircuitStructure}. The architecture is designed to assemble heterogeneous second-life battery packs directly to create a large-scale energy storage system. Other than the packs, it comprises DC/DC converters, a connection reconfigurator, and a power conversion system. Specifically, each pack is connected to a DC/DC converter that enables bidirectional power processing for charging/discharging power control of the pack. This pack-level power control capability crucially allows to handle the heterogeneity as well as operational safety of the packs. The connection configurator dynamically configures the connections between the packs (with the associated converters). It can choose to wire the packs in series, parallel or a mix of them to meet the DC link voltage requirements. It can also bypass the packs that are about to fail or reach the end of the second life to avoid system-level shutdown. The configurator thus provides the needed flexibility and reliability in reusing the heterogeneous second-life packs. Through the configurator and a shared DC link, the packs' charging/discharging power will be routed to the power conversion system, which interacts with the utility grid. Meanwhile, a power management system is in place to monitor and optimize the overall operation of the SL-BESS. Our prior study in \cite{TTE-FA-2023, 2021-IECON-FA} presents a pilot realization of the architecture, which leverages a circuit of power electronics switches as a connection configurator and DC/DC converters to regulate the power of the packs. 

The SL-BESS architecture offers several major benefits. The foremost one is that it makes it possible to combine and reuse heterogeneous second-life battery packs directly without refurbishment. This obviates the need to disassemble the packs and recondition and repackage the cells into new modules. By design, it also well allows to combine heterogeneous packs. The architecture, along with the power management system, can coordinate the power allocation from pack to system level. This paves the way for economic power management with an awareness of not only the external power demands but also the conditions of individual battery packs.

\subsection{Economic Performance Criteria}

\begin{figure}[!t]
\centering
\includegraphics[width=\linewidth]{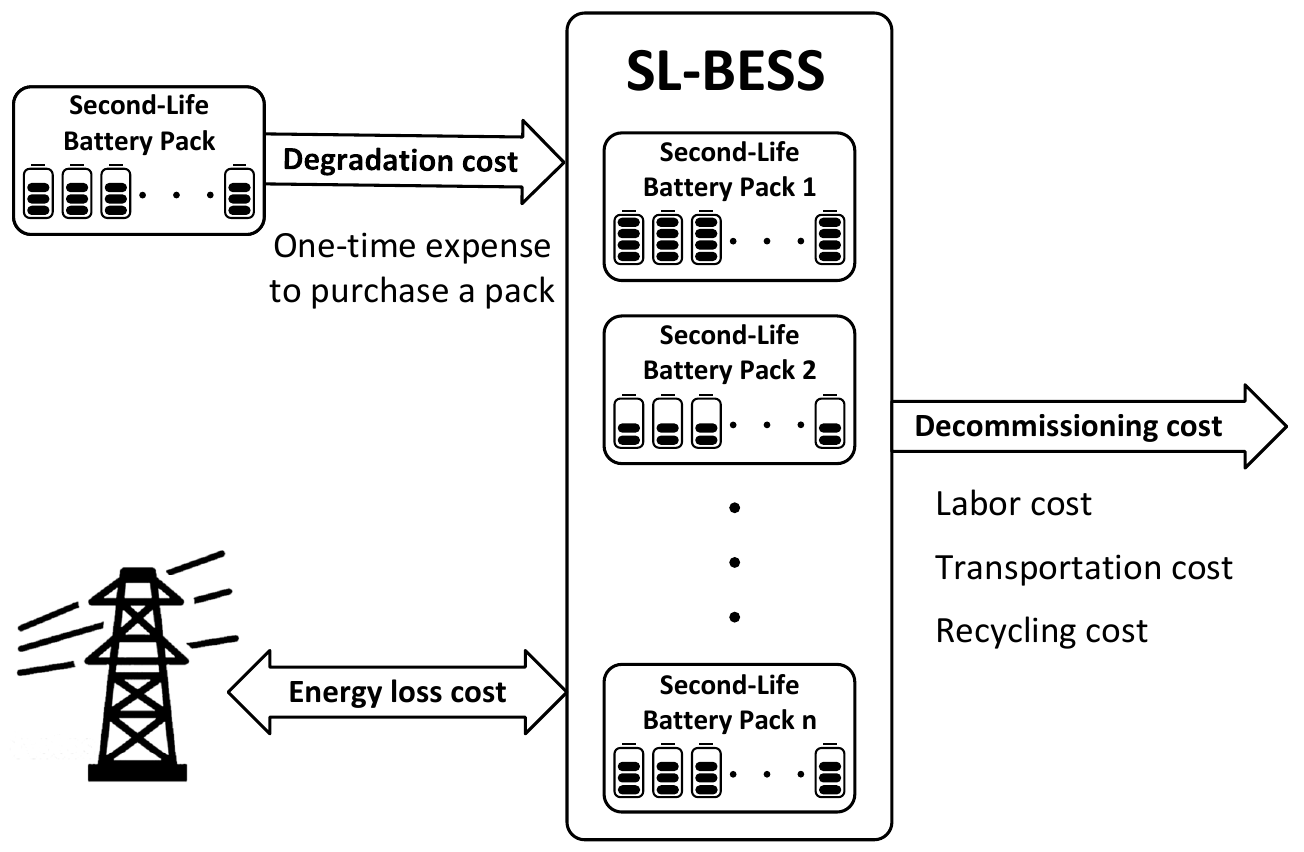}
\caption{The SL-BESS operating cost structure.}
\label{Cost}
\end{figure}

For the SL-BESS, the goal of power management is to maximize its economic performance. To this end, we consider the following costs associated with its operation, as depicted in Fig.~\ref{Cost}.

\begin{itemize}
	\item \textit{Degradation cost:} Second-life battery packs will degrade with cycled charging/discharging. The degradation is the loss of the remaining life, which is a long-term process subject to the nonlinear effects of the charging/discharging C-rates and operating temperatures. The resulting cost constitutes an important part of the overall cost structure of the SL-BESS, and is also associated with the distribution of the capital cost (the one-time expense to purchase a pack) over the second life depending on operating conditions.
	\item \textit{Energy loss cost:} The SL-BESS interacts with the grid to supply or draw power. The interaction involves power loss due to the charging/discharging inefficiency. This, together with the instantaneous electricity price, will incur the energy loss cost.
	\item \textit{Decommissioning cost:} At the end of the second life, the battery packs must be decommissioned. The decommissioning process will involve costs due to labor, transportation to the recycling facility, recycling, and possibly others. 
\end{itemize}

We emphasize that enforcing balancing strategies like SoC balancing among heterogeneous SL-BESS is neither practical nor beneficial. Instead, we aim for a cost-aware power management approach that respects and leverages these heterogeneities to achieve economic efficiency. For instance, as shown in Figs.~\ref{CircuitStructure}-\ref{Cost}, while SoC values differ across various SL-BESS, the SoC of the constituent cells within each individual SL-BESS is balanced by its battery management system (BMS). This paper focuses on economic power management among different SL-BESS, with the individual BMS being beyond its scope.

Next, we will derive the mathematical models to capture the above costs and use them to enable economic optimal power management. The proposed economic optimal power management approach aims to reach economically optimal operation, and at the same time, ensure the safe operation of the second-life battery packs while meeting the output power demand.

\section{Economic Optimal Power Management}
In this section, we will first develop the operating cost models dependent on the electrical, degradation, and thermal dynamics of second-life battery packs, and then present the economic optimal power management approach.
 
\subsection{Electrical Modeling}

Suppose that the SL-BESS has $n$ second-life battery packs. The energy dynamics of the $j$-th pack is governed by
\begin{equation}
	E_j[k+1] = E_j[k] + \left(P_{j,\textrm{ch}}[k]\eta_{j,\textrm{ch}} - \frac{P_{j,\textrm{disch}}[k]}{\eta_{j,\textrm{disch}}}\right)\Delta t,
	\label{EnergyDynamics}
\end{equation}
where $E_j$, $P_{j, \textrm{ch}} \geq 0$, and $P_{j, \textrm{disch}} \geq 0$ are the remaining energy, charging, and discharging power, respectively, $\eta_{j, \textrm{ch}},\eta_{j, \textrm{disch}}\in(0,1)$ represent the round-trip efficiency (RTE) in charging/discharging, respectively, and $\Delta t$ is the time step \cite{NetworkSystems-YT-2019}. The following constraint ensures that the battery pack is either in the charging or discharging mode:
\begin{equation}
	P_{j,\textrm{ch}}P_{j,\textrm{disch}} = 0.
\end{equation}
We further impose limits on $P_{j, \textrm{ch/disch}}$ and $E_j$ to guarantee the safety of the packs:
\begin{subequations}
	\begin{align}
		P_{j,\textrm{ch/disch}}^{\textrm{min}} &\leq P_{j,\textrm{ch/disch}} \leq P_{j,\textrm{ch/disch}}^{\textrm{max}}, \\
		E_j^{\textrm{min}} &\leq E_j \leq E_j^{\textrm{max}},
	\end{align}
	\label{SafetyConstraints}
\end{subequations}
\hspace{-5pt}where $P_{j, \textrm{ch/disch}}^{\textrm{min/max}}$ and $E_j^{\textrm{min/max}}$ denote the lower/upper safety bounds, respectively. The system-level supply-demand balance is also maintained by the following constraint
\begin{equation}
	\sum_{j=1}^n \left(P_{j,\textrm{disch}} - P_{j,\textrm{ch}}\right) = P_{\textrm{out}},
\end{equation}
where $P_{\textrm{out}}$ is the external power demand. Note that $P_{\textrm{out}} > 0$ indicates discharging, and $P_{\textrm{out}} < 0$ means charging.

During the charging and discharging cycles, SL-BESS experiences power dissipation due to the inefficiencies. Based on \eqref{EnergyDynamics}, the power loss $P_{j,\textrm{loss}}$ is given by
\begin{equation}
	P_{j,\textrm{loss}}[k] =	P_{j,\textrm{ch}}[k](1-\eta_{j,\textrm{ch}}) + P_{j,\textrm{disch}}[k]\left(\frac{1}{\eta_{j,\textrm{disch}}}-1\right).
	\label{PowerLoss}
\end{equation}
\subsection{Degradation Modeling}
EV batteries degrade over time in repeated charging and charging cycles, as a result of complex parasitic or side reactions and influenced by driving patterns and ambient temperature in the first life \cite{SAE-NJ-2015, TTE-VW-2022}. The degree of degradation is often quantified using SoH:
\begin{equation}
	\textrm{SoH}_j = \frac{Q_j - Q_{j,\textrm{fade}}}{Q_j},
	\nonumber
	\label{SoH}
\end{equation}
where $Q_{j,\textrm{fade}}$ and $Q_j$ are the capacity fade and the initial capacity when the pack is new, respectively. Capacity fading modeling of lithium-ion batteries has attracted significant research efforts \cite{TSG-LS-2024, 
PowerSources-JZ-2021}. Physics-based electrochemical models deliver accurate predictions but are too complex to be useful for addressing economic optimization problems. An appealing alternative is the weighted Ah-throughput models, which associate the energy throughput with the loss of life \cite{SmartGrid-LM-2017}. These models have simpler mathematical structure and thus lend well to optimization, while allowing for relatively easier experimental calibration. In what follows, we introduce a weighted Ah-throughput model for the context of second-life battery packs.

The model begins with an underlying assumption that a battery pack can deliver a specific Ah-throughput under nominal operating conditions defined in terms of C-rate and temperature before the end of life \cite{InternationalPowerElectronics-OS-2012}. When the actual operating conditions deviate from the nominal ones, the physical Ah-throughput that the pack gives will increase or decrease, impacting the degradation of the pack differently. We present the following weighted Ah-throughput model for the $j$-th battery pack, which is from \cite{PowerSources-JW-2011}:
\begin{equation}
	Q_{j,\textrm{fade}} = B\exp\left({\frac{-(E + \beta C_{j})}{RT_j}}\right)\mathcal{Z}_j^\zeta.
	\label{Qfade}
\end{equation}
Here, $Q_{j,\textrm{fade}}$, $\mathcal{Z}_j$, $T_j$, and $C_j$ are the loss of capacity, consumed Ah-throughput, pack's temperature, and C-rate, respectively. Further, $E$ and $R$ are the activation energy and gas constant, respectively; $B$, $\beta$, and $\zeta$ are the pre-exponential, compensation, and power law factors, respectively. The model quantifies the capacity fading when a given Ah-throughput is extracted from the pack under a specified temperature and C-rate. It shows the weighting effect of the operating conditions on the degradation, in reflection of the model's name.

By \eqref{Qfade}, we have
\begin{equation}
	\frac{dQ_{j,\textrm{fade}}}{d\mathcal{Z}_j} = B\zeta\exp\left({\frac{-(E + \beta C_{j})}{RT_j}}\right)\mathcal{Z}_j^{\zeta - 1}.
	\label{Derivative}
\end{equation}
We approximate the derivative in \eqref{Derivative} using the forward Euler method as follows:
\begin{equation}
	\begin{split}
		\frac{dQ_{j,\textrm{fade}}}{d\mathcal{Z}_j} \approx & \; \frac{Q_{j,\textrm{fade}}[k+1] - Q_{j,\textrm{fade}}[k]}{\mathcal{Z}_j[k+1]-\mathcal{Z}_j[k]} = \\
		&B\zeta\exp\left({\frac{-(E + \beta C_{j})}{RT_j}}\right)\mathcal{Z}_j[k]^{\zeta - 1}.
	\end{split}
	\label{ApproxDerivative}
\end{equation}
It follows that
\begin{equation}
	\begin{split}
	Q_{j,\textrm{fade}}&[k+1] = Q_{j,\textrm{fade}}[k] + \\
	& \Delta \mathcal{Z}_j[k]B^{\frac{1}{\zeta}}\zeta\exp\left({\frac{-(E + \beta C_{j})}{\zeta RT_j}}\right)Q_{j,\textrm{fade}}^{\frac{\zeta - 1}{\zeta}}[k],
	\end{split}
	\label{AgingModel}
\end{equation}
where $\Delta \mathcal{Z}_j$ can be calculated by
\begin{equation}
	\Delta \mathcal{Z}_j[k] = \frac{1}{3600}\frac{P_{j,\textrm{ch/disch}}[k]}{V_j} \Delta t,
	\label{Ah}
\end{equation}
where $V_j$ and $P_{j,\textrm{ch/disch}}$ are the nominal voltage and charging/discharging power of the battery pack, respectively. The relationships shown by \eqref{AgingModel}-\eqref{Ah} capture how the charging/discharging power affects the capacity fading. Its concise structure conveys computational tractability as will be needed later in economic power optimization.  

However, there is a remaining issue---\eqref{AgingModel} is temperature-dependent and thus requires a model to grasp the effect of the charging/discharging power $P_{j,\textrm{ch/disch}}$ on the pack's temperature $T_j$. Next, we address this issue. 

\subsection{Thermal Modeling}
We employ a lumped thermal model to characterize the temperature of the second-life battery packs \cite{2016-TSTE-PC}. This model considers the internal power loss and convection as the primary sources of heat generation and dissipation, respectively. Its governing equation is as follows:
\begin{equation}
	m_jc_{j,\textrm{th}}\dot{T_j}(t) = R_0i^2 - (T_j(t)-T_{\textrm{env}})/R_{j,\textrm{conv}},
	\label{CellLevelThermalModel}
\end{equation}
where $T_j$ and $T_{\textrm{env}}$ are the pack's aggregate temperature and the environmental temperature, respectively. Further, $m_j$ and $c_{j,\textrm{th}}$ are the mass and specific heat capacity, respectively; the power loss is captured by $R_0i^2$, where $R_0$ and $i$ are the equivalent internal resistance and current of the battery pack, respectively; $R_{j,\textrm{conv}}$ is the convective thermal resistance between the battery pack and the environment. More specifically, $R_{j,\textrm{conv}}$ can be specified by 
\begin{equation}
	R_{j,\textrm{conv}} = \frac{1}{hA_j},
	\nonumber
\end{equation}
where $h$ and $A_j$ are the heat transfer coefficient between the pack and environment, and the external surface area of the pack, respectively. 

We proceed further to consider the steady-state behavior of the temperature under different C-rates. Given $T_{\textrm{env}}$ and leveraging \eqref{CellLevelThermalModel}, we can find out the relationship between $C_{j}$ and $T_j$, which can be approximated by a second-order polynomial: 
\begin{equation}
	T_j = \alpha_{j,1} + \alpha_{j,2}C_{j} + \alpha_{j,3}C_{j}^2,
	\label{Temp}
\end{equation}
where $\alpha_{j,i}, i=1,2,3$ are the polynomial's coefficients. Note that $\alpha_i$ varies for different $T_{\textrm{env}}$, and that $C_{j}$ can be expressed in terms of $P_{j,\textrm{ch/disch}}$ using $C_{j}=P_{j,\textrm{ch/disch}}/V_jQ_j$. Later, we will use the relationship in \eqref{Temp}, rather than \eqref{CellLevelThermalModel}, in economic optimization for power management to reduce the computational complexity.

\begin{figure}[!t]
\centering
\includegraphics[trim={2.4cm 0.5 2.8cm 1cm}, clip, width=\linewidth]{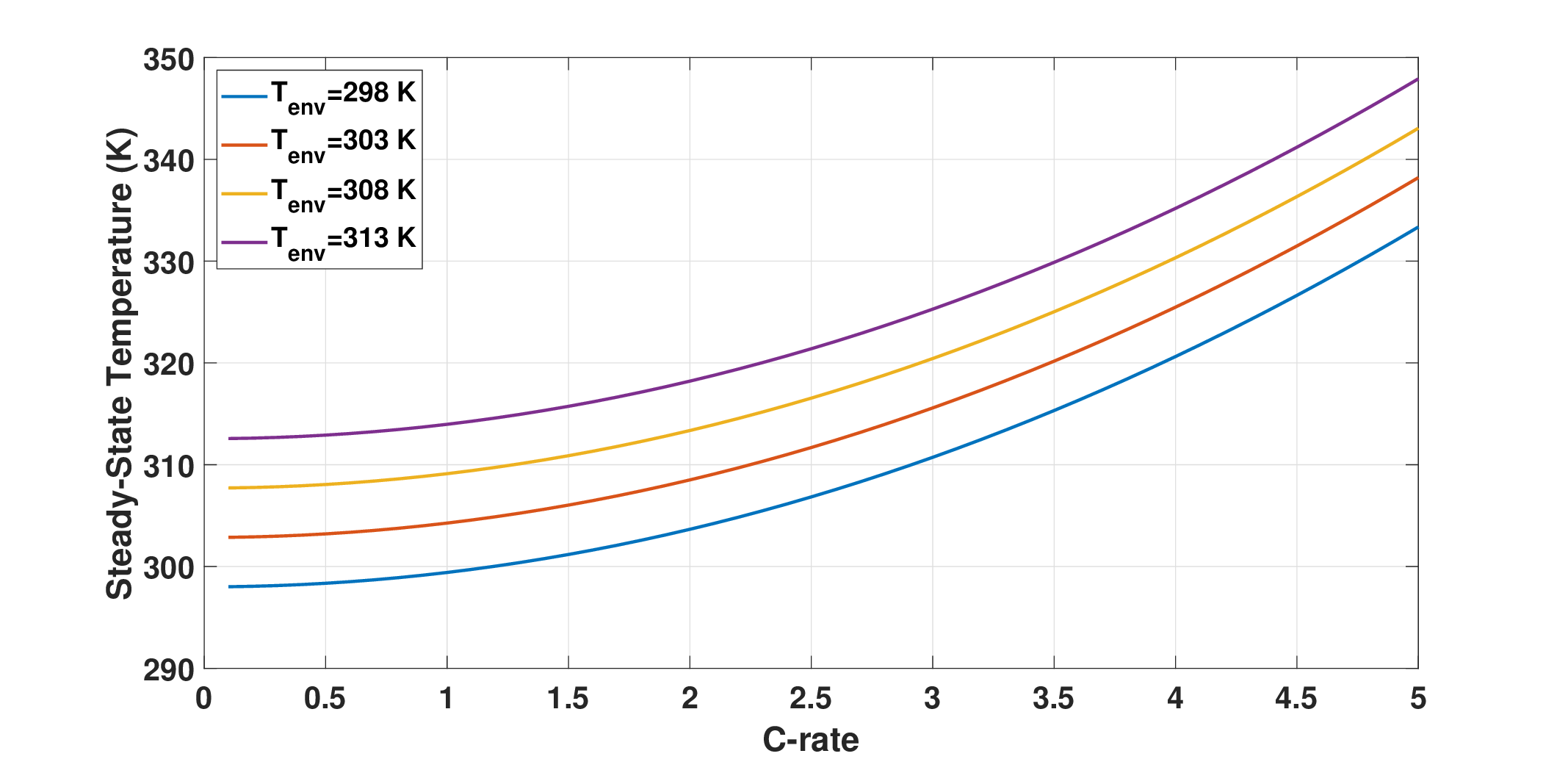}
\caption{Steady-state temperature of the battery pack under varying C-rates and environmental temperatures. The constituting cells' parameters are $h=5.8$, $m=0.072$ kg, $c_{\textrm{th}}=1150$ J/kg.K, $A=0.0053$ m\textsuperscript{2}, $R_0=0.008$ $\Omega$.}
\label{SS_Temp}
\end{figure}
\subsection{A Case Study}

\begin{table}[!t]
	\renewcommand{\arraystretch}{1.2}
	\caption{Aging Model Parameters}
	\centering
	\label{TABLE}
	\resizebox{\columnwidth}{!}{
		\begin{tabular}{l l l}
			\hline\hline \\[-3mm]
			\multicolumn{1}{c}{Symbol} & \multicolumn{1}{c}{Parameter} & \multicolumn{1}{c}{Value [Unit]}  \\[1.6ex] \hline
			$ Q $  & Capacity & 60 kWh \\
			$ V $  & Nominal voltage & 380 V \\
			$ E $ & Activation energy & 31700 J \\
			$ \beta $ & C-rate compensation factor & -370.3 \\ 
			$ \zeta $ & Power law factor & 0.55 \\
			$ R $ & Gas constant & 8.31 J/(mol.K) \\ 
			$ T_{\textrm{env}} $ & Environment temperature & 298 K \\
 			$ \alpha_i, i=1,2,3 $ & Coefficients in \eqref{Temp} & 298, 0, 1.421 \\
			$ \Delta t $ & Sampling time & 1 hour\\
			\hline\hline
		\end{tabular}
	}
\end{table}

Leveraging the models in Sections III.A-C, we present a case study to show the capacity fading of a second-life battery pack under varying operating conditions. The pack consists of 7104 A123 ANR26650M1-A LiFePO4 cylindrical cells, which are configured as 16 serially connected 6S74P modules. Its total power is 60 kWh with a nominal voltage of 380 V. The pack has lost 20\% of its original capacity after the first life, and is allowed to consume another 10\% of its capacity in the second life. Table~\ref{TABLE} summarizes the parameters of the capacity fading model. The pre-exponential factor $B$ depends on C-rate and is calculated as follows:
\begin{equation}
	B_j = 3172.4 - 590.66C_j + 42.08C_j^2.
	\label{B12}
\end{equation}
Running the models, we obtain a catalog of results. 

\begin{figure}[!t]
\centering
\includegraphics[trim={2.8cm 0.5 2.8cm 1cm}, clip, width=\linewidth]{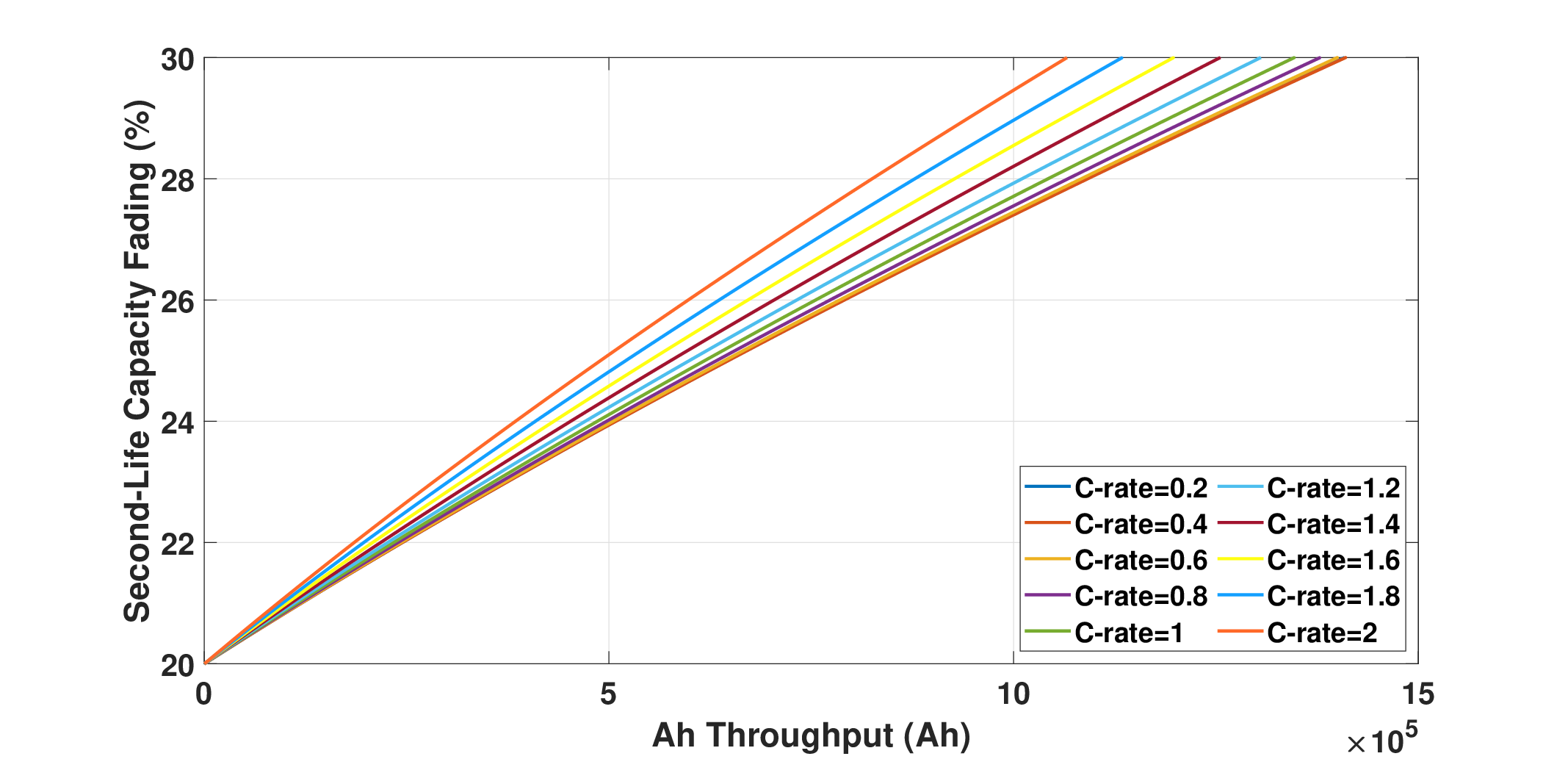}
\caption{Second-life capacity fading in relation to Ah-throughput.}
\label{Q_fade}
\end{figure}

\begin{figure}[!t]
\centering
\includegraphics[trim={2.8cm 0.5 2.8cm 0.4cm}, clip, width=\linewidth]{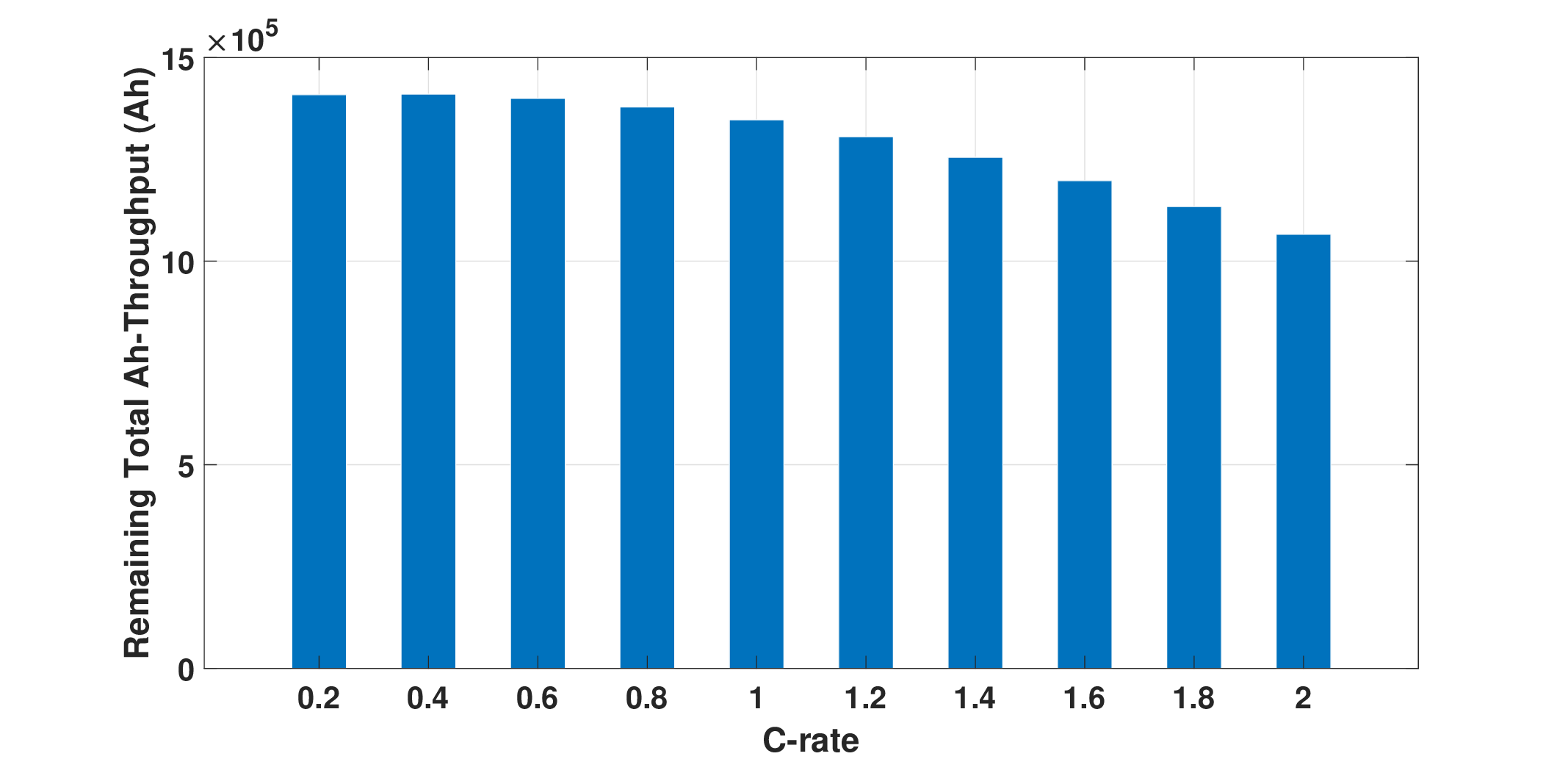}
\caption{The remaining total Ah-throughput with respect to C-rate.}
\label{BarAh}
\end{figure}

Fig.~\ref{SS_Temp} depicts the steady-state temperature profiles versus C-rates under different $T_{\textrm{env}}$. Further, Fig.~\ref{Q_fade} illustrates the cumulative capacity fading in the second life for different C-rates, and Fig.~\ref{BarAh} shows the usable total Ah-throughput versus different C-rates. The results suggest that the capacity fading is a nonlinear process subject to the effects of C-rates as well as the thermal dynamics. Especially, higher C-rates accelerate the aging and degradation while reducing the Ah-throughput that can be extracted in the second life. These impacts have significant economic implications to require economy-conscious power management.

\subsection{Cost Modeling}

The power loss results in an economic cost that depends on the instantaneous electricity price. Considering an SL-BESS with $n$ battery packs, the cumulative energy loss cost $C_{\textrm{RTE}}$ within the planning horizon $[0, H]$ is given by
\begin{equation}
	C_{\textrm{RTE}} = \sum_{k=0}^H\sum_{j=1}^n \tau[k]P_{j,\textrm{loss}}[k],
\end{equation}
where $\tau[k]$ is the electricity price at time $k$. We highlight that $\eta_{j,\textrm{ch/disch}}$ vary across the second-life battery packs due to discrepancies in their chemistry, capacity, and aging condition. This makes $C_{\textrm{RTE}}$ an important factor to be considered in power allocation among individual packs for optimal economic performance.

The degradation of a battery pack incurs degradation and decommissioning costs. With the developed degradation model, we can now express these costs associated with the capacity fading. The total degradation cost within the planning horizon $[0, H]$ is given by
\begin{equation}
	C_{\textrm{deg}} = \sum_{j=1}^{n} \frac{C_{j, \textrm{cap}}}{Q_{j,\textrm{SL}}}\left(Q_{j,\textrm{fade}}[H]-Q_{j,\textrm{fade}}[0]\right),
	\label{CapitalCost}
\end{equation} 
where $C_{\textrm{deg}}$, $C_{j, \textrm{cap}}$, and $ Q_{j,\textrm{SL}}$ are the overall degradation cost, capital (purchase) cost and remaining useful second life, respectively. The expression in \eqref{CapitalCost} calculates the degradation cost as the sum of the individual contributions from each battery pack, determined by the difference between the capacity fading at the end and the beginning of the planning horizon.

The decommissioning cost encompasses the costs for labor, transportation, recycling, and others to remove a battery pack from service at the end of its second life. It is typically expressed in terms of dollars per pound due to the dependence on the mass \cite{PNNL-WV-2022}. We denote the per-unit-mass decommissioning cost as $\bar{C}_{\textrm{decom}}$. Even though it is a one-time expense to decommission a battery pack, the cost manifests itself in every operating intervals. This is because the actual length of the second life relies on the operating conditions as revealed by the degradation model in Section III.B. Then, the total decommissioning cost for the SL-BESS within the planning horizon $[0,H]$ is given by
\begin{equation}
	C_{\textrm{decom}} = \sum_{j=1}^{n} \frac{m_j\bar{C}_{\textrm{decom}}}{Q_{j,\textrm{SL}}}\left(Q_{j,\textrm{fade}}[H]-Q_{j,\textrm{fade}}[0]\right).
	\label{DecommissionCost}
\end{equation} 
Note that \eqref{CapitalCost} and \eqref{DecommissionCost} effectively distribute the long-term economic costs for each planning horizon. This enables us to reconcile the short-term energy loss cost with the long-term capital and decommissioning costs in the problem formulation. 

\subsection{Problem Formulation}

\renewcommand{\baselinestretch}{1}

With the above cost models, we are now ready to consider economic optimization in power management for the SL-BESS. To begin with, we gather the optimization variables in a vector $z_j$, where $z_j=\begin{bmatrix} P_{j,\textrm{ch/disch}} & E_j & Q_{j, \textrm{fade}} \end{bmatrix}^\top, j=1,…,n$. We propose an approach for economic optimal power management by addressing the following constrained optimization problem: 
\vspace{-5pt}
\begin{equation}
	\begin{aligned}
		&\min_{z_j, j=1,...,n} C_{\textrm{RTE}} + C_{\textrm{deg}} + C_{\textrm{decom}},\\ 
		&\textrm{Safety constraints:} \quad \\
		&\quad P_{j,\textrm{ch/disch}}^{\textrm{min}} \leq P_{j,\textrm{ch/disch}} \leq P_{j,\textrm{ch/disch}}^{\textrm{max}},\\
		&\quad E_j^{\textrm{min}} \leq E_j \leq E_j^{\textrm{max}},\\
		&\quad P_{j,\textrm{ch}}P_{j,\textrm{disch}} = 0,\\
		&\textrm{Energy dynamics:} \quad \\
		&\quad E_j[k+1] = E_j[k] + \left(P_{j,\textrm{ch}}[k]\eta_{j,\textrm{ch}} - \frac{P_{j,\textrm{disch}}[k]}{\eta_{j,\textrm{disch}}}\right)\Delta t, \\
		&\textrm{Aging dynamics:} \quad \\
		&\quad Q_{j,\textrm{fade}}[k+1] = Q_{j,\textrm{fade}}[k] + \\
		& \qquad \qquad \Delta \mathcal{Z}_j[k]B_j^{\frac{1}{\zeta_j}}\zeta_j\exp\left({\frac{-(E_j + \beta_j C_j)}{\zeta_j RT_j}}\right)Q_{j,\textrm{fade}}^{\frac{\zeta_j - 1}{\zeta_j}}[k],\\
		&\textrm{Temperature dynamics:} \quad \\
		& \quad T_j = \alpha_{j,1} + \alpha_{j,2}C_{j} + \alpha_{j,3}C_{j}^2,\\
		&\textrm{Power supply-demand balance:} \quad \\
		&\quad \sum_{j=1}^n \left(P_{j,\textrm{disch}} - P_{j,\textrm{ch}}\right) = P_{\textrm{out}}.
		\label{Optim}
	\end{aligned}
\end{equation}
The proposed approach focuses on maximizing the economic performance of SL-BESS in predictive power management. It takes into account the heterogeneity among the battery packs within the SL-BESS to allocate the power while adhering to the constraints ensuring safety and power-demand balance. Also note that we do not impose SoC and SoH balancing constraints among the second-life battery packs in \eqref{Optim} as their benefits to the SL-BESS are unclear. However, we acknowledge battery pack heterogeneity by allowing flexible SoC and SoH levels to maximize economic benefits. The approach is the first one that we are aware of that deals with economic optimal power management for SL-BESS, which can find prospective use in different SL-BESS applications in smart grid, renewable energy, among others. 
\section{Simulation Results}
\begin{table}[!t]
	\renewcommand{\arraystretch}{1.3}
	\caption{Second-Life Battery Pack Specifications}
	\centering
	\label{TABLE-SIM-1}
	\resizebox{\columnwidth}{!}{
		\begin{tabular}{c l l l l l l l}
			\hline\hline \\[-3mm]
			\multicolumn{1}{c}{Type} & \multicolumn{1}{c}{Battery packs} & \multicolumn{1}{c}{Capacity} & \multicolumn{1}{c}{$\eta_{\textrm{ch/disch}}$} & \multicolumn{1}{c}{Capital cost} & \multicolumn{1}{c}{$\textrm{SoH}$} & \multicolumn{1}{c}{$Q_{\textrm{SL}}$}\\[1.6ex] \hline
			$ 1 $  & 1-20   & 60 kWh & 85\% & 90 \$/kWh & 85\% & 15\%   \\
			$ 2 $  & 21-40 & 60 kWh & 80\% & 75 \$/kWh & 80\% & 10\%   \\
			$ 3 $  & 41-60 & 40 kWh & 85\% & 100 \$/kWh & 85\%  & 15\%   \\
			$ 4 $  & 61-80 & 40 kWh & 80\% & 80 \$/kWh & 80\% & 10\%   \\
			\hline\hline
		\end{tabular}
	}
\end{table}

\begin{table}[!t]
	\renewcommand{\arraystretch}{1.3}
	\caption{Safety and Cost Specifications of the Considered SL-BESS}
	\centering
	\label{TABLE-SIM-2}
	\resizebox{\columnwidth}{!}{
		\begin{tabular}{l l l}
			\hline\hline \\[-3mm]
			\multicolumn{1}{c}{Symbol} & \multicolumn{1}{c}{Parameter} & \multicolumn{1}{c}{Value}  \\[1.6ex] \hline
			$ m $ & Per unit mass & 4.42 Pounds/kWh \\
			
			$ \bar{C}_{\textrm{decom}} $ 	& Per-unit-mass decommissioning cost			& 1.75 \$/Pound  \\

			$ E_j^{\textrm{min}} $ & Minimum remaining energy & $0.15Q_j$ kWh \\
			$ E_j^{\textrm{max}} $ & Maximum remaining energy & $0.85Q_j$ kWh \\
			
			$ P_{j,\textrm{ch,disch}}^{\textrm{min}} $ & Minimum charging/discharging power & $0.05Q_j$ kW \\
			$ P_{j,\textrm{ch,disch}}^{\textrm{max}} $ & Maximum charging/discharging power & $0.5Q_j$ kW\\

			\hline\hline
		\end{tabular}
	}
\end{table}

This section aims to validate the efficacy of the proposed economic optimal power management approach through simulations. Here, we examine a 2 MWh SL-BESS comprising 80 second-life battery packs. These packs fall into four types in terms of capacity, RTE, purchase price, and remaining useful second life, as summarized in Table \ref{TABLE-SIM-1}. The heterogeneity among them will allow us to examine how this influences the power optimization. Further, Table \ref{TABLE-SIM-2} details the operating ranges and cost specifications, where the maximum/minimum charging/discharging power and remaining energy level of each battery pack are set relative to its capacity. We examine two cases for the aging dynamics parameters. 
\begin{itemize}
	\item {\bf Case 1} uses the second-life A123 ANR26650M1-A LiFePO4 cylindrical cells in \cite{PowerSources-JW-2011}, with the cell specifications detailed in Table \ref{TABLE};
	\item {\bf Case 2} uses the prismatic LiFePO4 cells in \cite{AppliedEnergy-ZS-2014}, which have a faster pace degradation than the cells in Case 1. 
\end{itemize}

\begin{figure}[!t]
\centering
\includegraphics[trim={2.2cm 0.5 2.8cm 0.4cm}, clip, width=8cm]{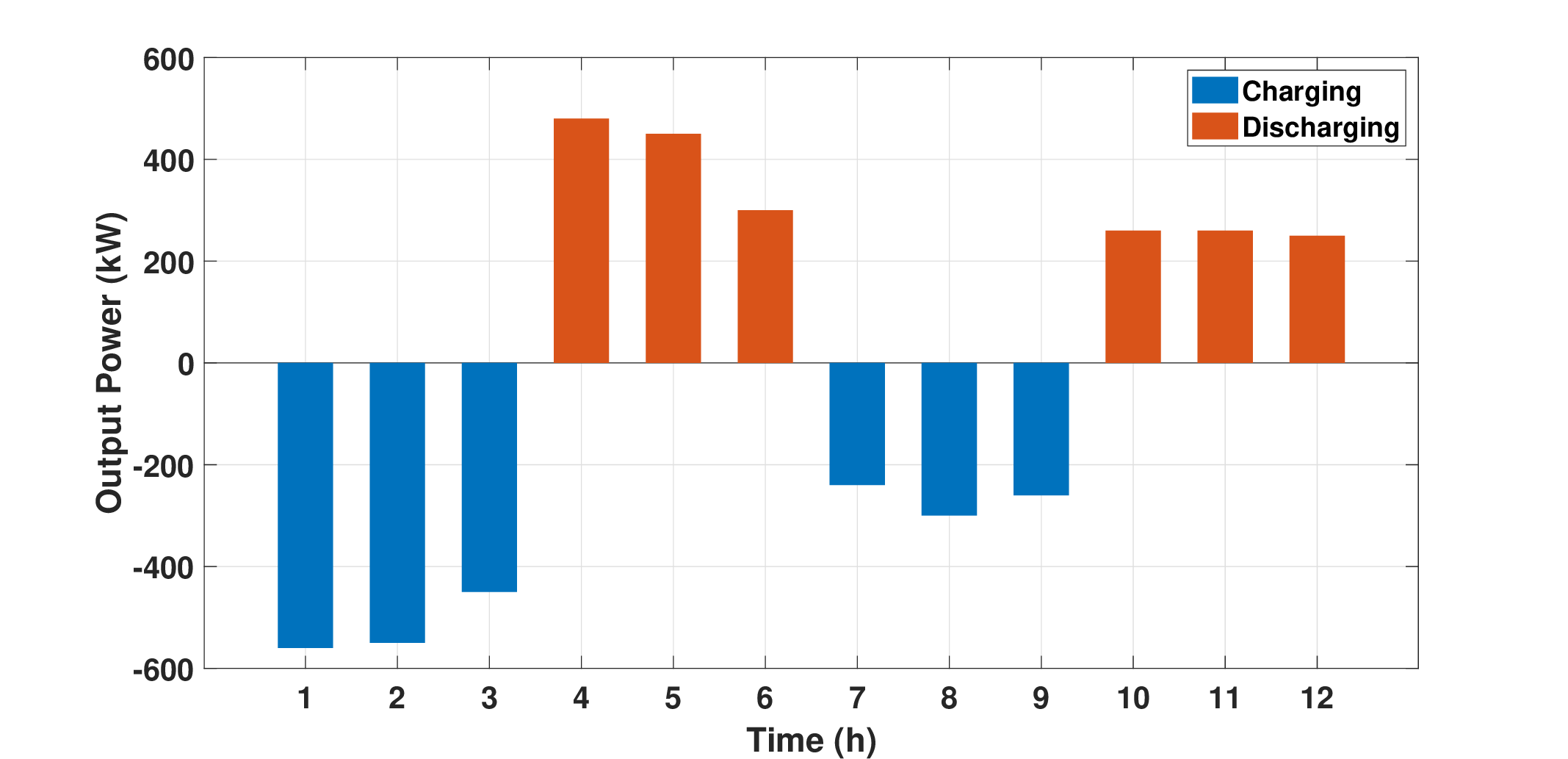}
\caption{The output power profile of the SL-BESS.}
\label{Pout}
\end{figure}

The SL-BESS is set to supply given power within a planning horizon of 12 hours, with the power demand profile shown in Fig.~\ref{Pout}. The electricity price changes hourly and for the purpose of evaluation, is stochastic to follow a Gaussian distribution with a mean of 0.15 \$/kWh and a covariance of 0.01 \$/kWh. The constituent packs all start with 20\% of SoC in the simulations. 

We will also compare the performance of the proposed economic optimal power management approach with two other approaches: SoH-based and capacity-based power allocation. These approaches mandate the power allocation to be governed, respectively, by
\begin{align}
	P^{\textrm{SoH}}_{j,\textrm{ch,disch}} =& \frac{\textrm{SoH}_j}{\sum_{i=1}^n \textrm{SoH}_i}P_{\textrm{out}}, \nonumber \\
	P^Q_{j,\textrm{ch,disch}} =& \frac{Q_j}{\sum_{i=1}^n Q_i}P_{\textrm{out}}. \nonumber
\end{align}

\subsection{Techno-Economic Analysis: Case 1}
In this case, the second-life battery packs comprise A123 ANR26650M1-A LiFePO4 cylindrical cells, and their parameters are detailed in Table \ref{TABLE}. However, the pre-exponential factor $B_j$ varies among different battery pack types. For battery packs of Type 1 and 2, the pre-exponential factor is given in \eqref{B12}, while for those of Type 3 and 4, it is as follows:
\begin{equation}	
	B_j = 3851.4 - 717.16C_j + 51.09C_j^2.
	\label{B34}
\end{equation}
We will repeatedly apply the output power profile depicted in Fig.~\ref{Pout} to the SL-BESS for 4800 cycles to assess the long-term aging of the second-life battery packs. 

\begin{figure}[!t]
\centering
\includegraphics[trim={0cm 0cm 0cm 0cm}, clip, width=8cm]{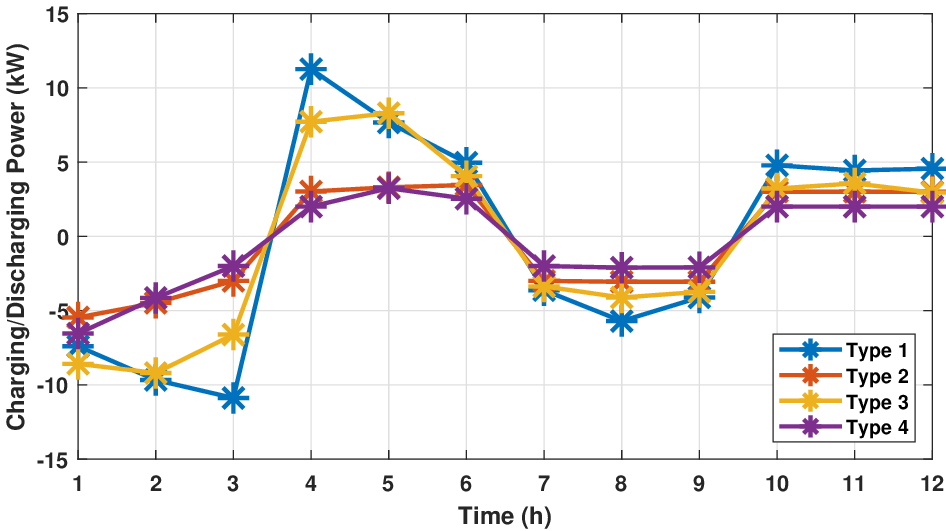}
\caption{The charging/discharging power profiles of the second-life battery packs.}
\label{Pch-dis}
\end{figure}

Fig.~\ref{Pch-dis} illustrates the charging/discharging power profiles at the 2400-th cycle, which correspond to the four types summarized in Table \ref{TABLE-SIM-1}. As is seen, the packs of Type 1 and Type 3 provide noticeably higher charging/discharging power compared to those of Types 2 and 4. This is because the packs of Type 1 and Type 3 have higher RTE and SoH when the second life begins. Comparing the packs of Type 1 and Type 3 with similar RTE and SoH, we observe that the packs of Type 3 have slightly lower charging/discharging power due to their marginally higher capital cost. It is important to notice that the Type-3 battery packs have the highest capital cost with 100\$/kWh, followed by Type-1 battery packs priced at 90\$/kWh. However, their higher RTE and longer second life offset the effect of the capital cost, making them used more for better economic performance. In other words, Type 1 arises as the most economical choice among these four types of battery packs. This finding highlights the importance of considering RTE and remaining useful second life in setting up and managing SL-BESS.

\begin{figure}[!t]
\centering
\includegraphics[trim={0cm 0cm 0cm 0cm}, clip, width=8cm]{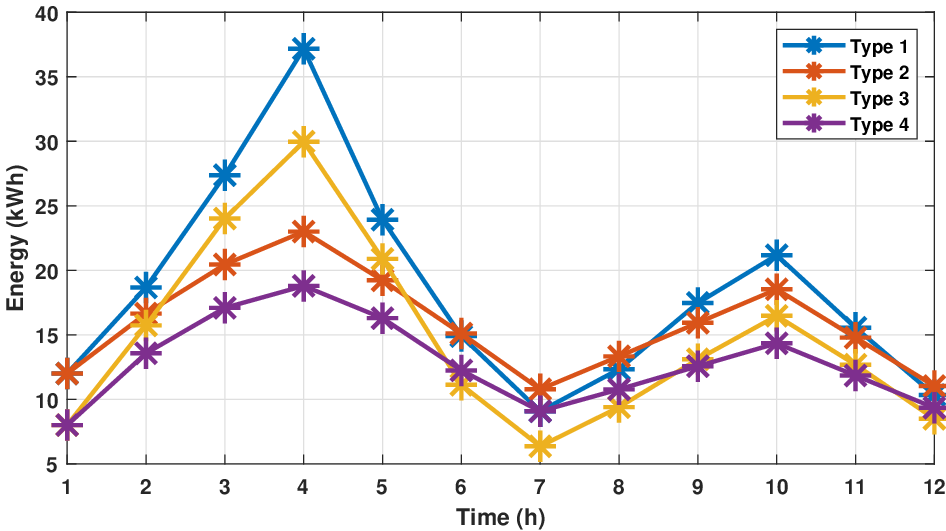}
\caption{The remaining energy of the second-life battery packs.}
\label{Energy}
\end{figure}

Fig.~\ref{Energy} depicts the remaining energy of the second-life battery packs. Consistent with Fig.~\ref{Pch-dis}, we see that the battery packs of Type 1 exhibit the most changes throughout the operation in their energy levels. It is in our interest to highlight that the energy levels of the battery packs are not balanced. As mentioned in Section I, balancing the energy level of heterogeneous second-life battery packs is neither useful nor necessary, as the packs are sourced from different vendors, present different characteristics, and have different first-life usage. A more practical goal is to optimize the economic performance while respecting the heterogeneity across the battery pack, as pursued in this study.

\begin{figure}[!t]
\centering
\includegraphics[trim={0cm 0cm 0cm 0cm}, clip, width=8cm]{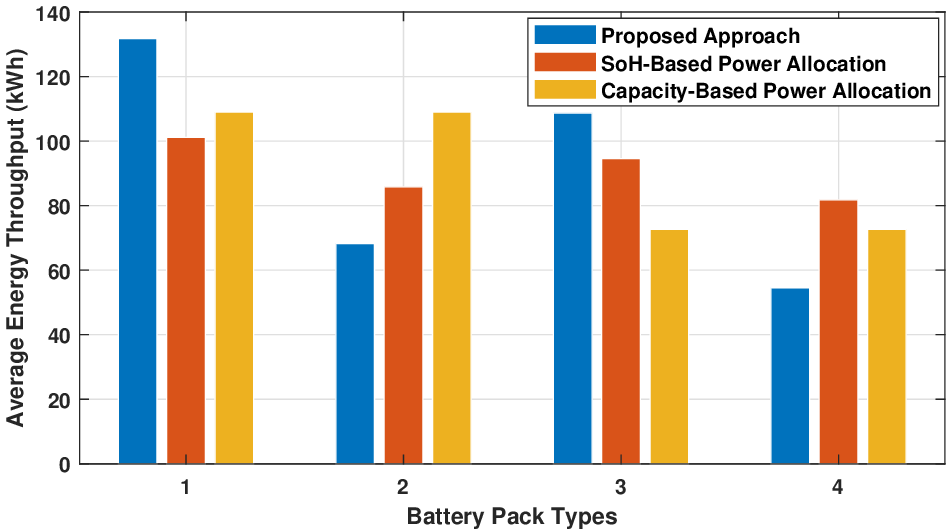}
\caption{The average energy throughput of second-life battery pack types.}
\label{Ah_AVG}
\end{figure}

Fig.~\ref{Ah_AVG} shows the average energy throughput of all battery pack types under the three power allocation approaches. With the SoH-based allocation, Types 1 and 3 are utilized more due to their higher SoH values, while Types 2 and 4 are less utilized. The capacity-based approach treats Types 1 and 2, and Types 3 and 4 equally. Fig.~\ref{SoH} also depicts the SoH of battery pack types over 4800 cycles. Our proposed approach utilizes the newer packs (Types 1 and 3) more, resulting in faster degradation for these types. Older packs (Types 2 and 4) are less utilized to extend their useful second life. Overall, the SoH-based allocation overlooks capacity differences, while the capacity-based allocation ignores SoH variations. In contrast, our approach leads to significantly different throughputs and SoH profiles, favoring the usage of Types 1 and 3 more, due to its focus on achieving better economic performance.

\begin{figure}[!t]
\centering
\includegraphics[trim={1cm 0.5cm 1cm 0.2cm}, clip, width=\linewidth]{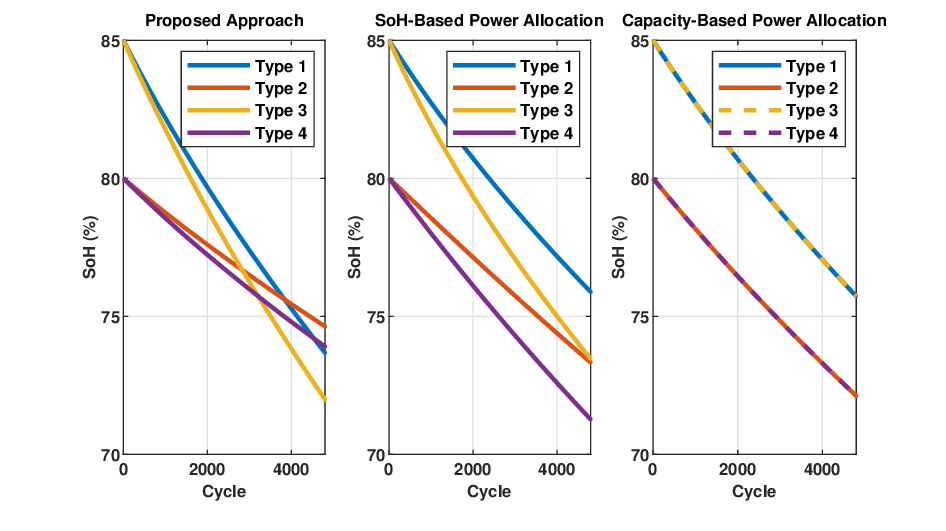}
\caption{The average degradation of different second-life battery pack types.}
\label{SoH}
\end{figure}

\begin{figure}[!t]
\centering
\includegraphics[trim={0.25cm 0.25cm 0.25cm 0.25cm},clip,width=8.5cm]{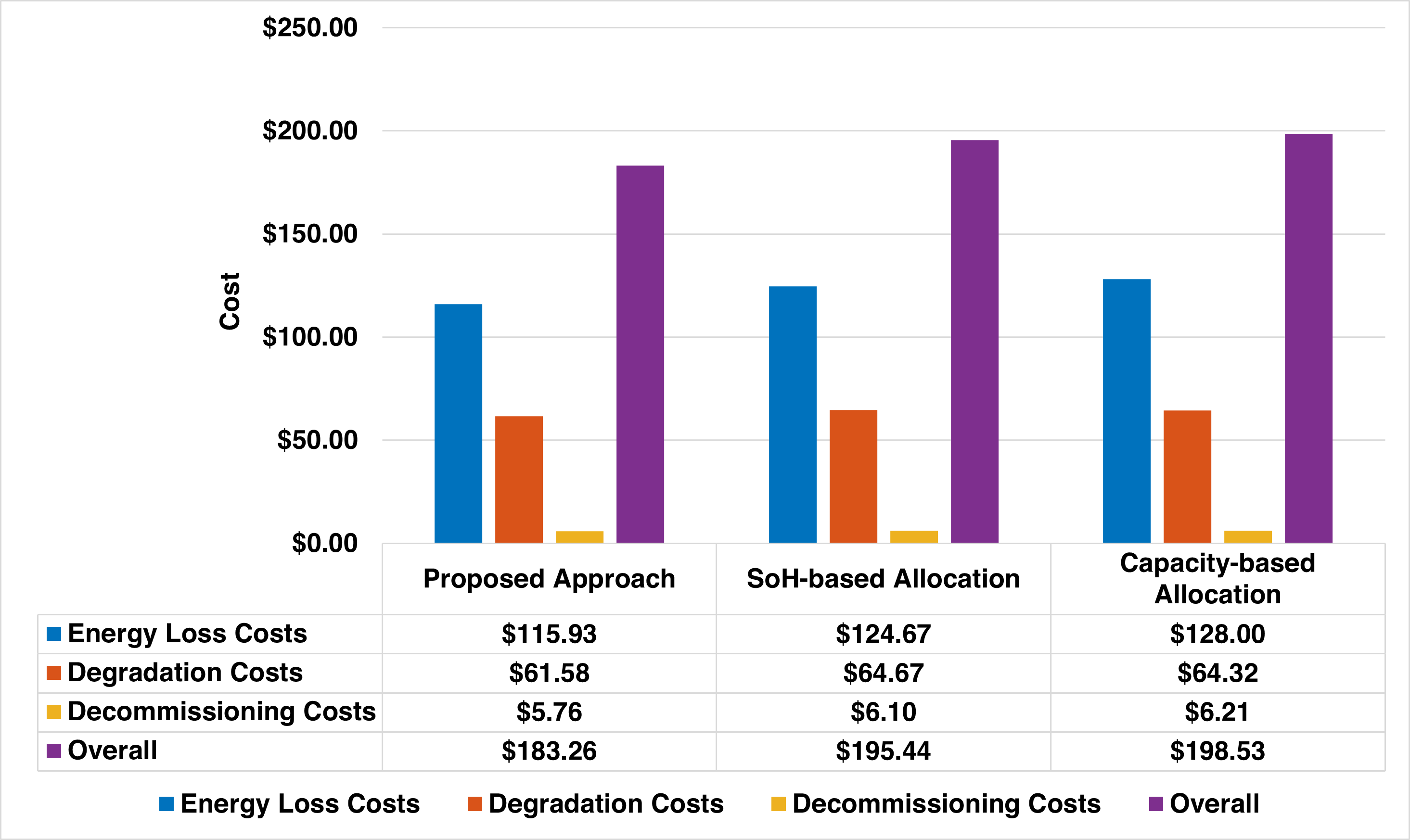} 
\caption{The cost analysis for different power allocation approaches for the first operation cycle.}
\label{Case1_1}
\end{figure}

Fig.~\ref{Case1_1} illustrates the total costs across various power allocation approaches for the first operation cycle as an example. The proposed approach achieves the lowest cost at \$183.26 with a reduction of approximately 6.23\% and 7.69\% compared to the SoH-based and capacity-based approaches, respectively. We also see that the energy loss cost is the most dominant cost, followed by the degradation cost. Further, decommissioning costs, while accounting for only 3-4\% of the overall cost, are still worthy to be considered for cost-effective operation.

\begin{figure}[!t]
\centering
\includegraphics[trim={0.25cm 0.25cm 0.25cm 0.25cm},clip,width=8.5cm]{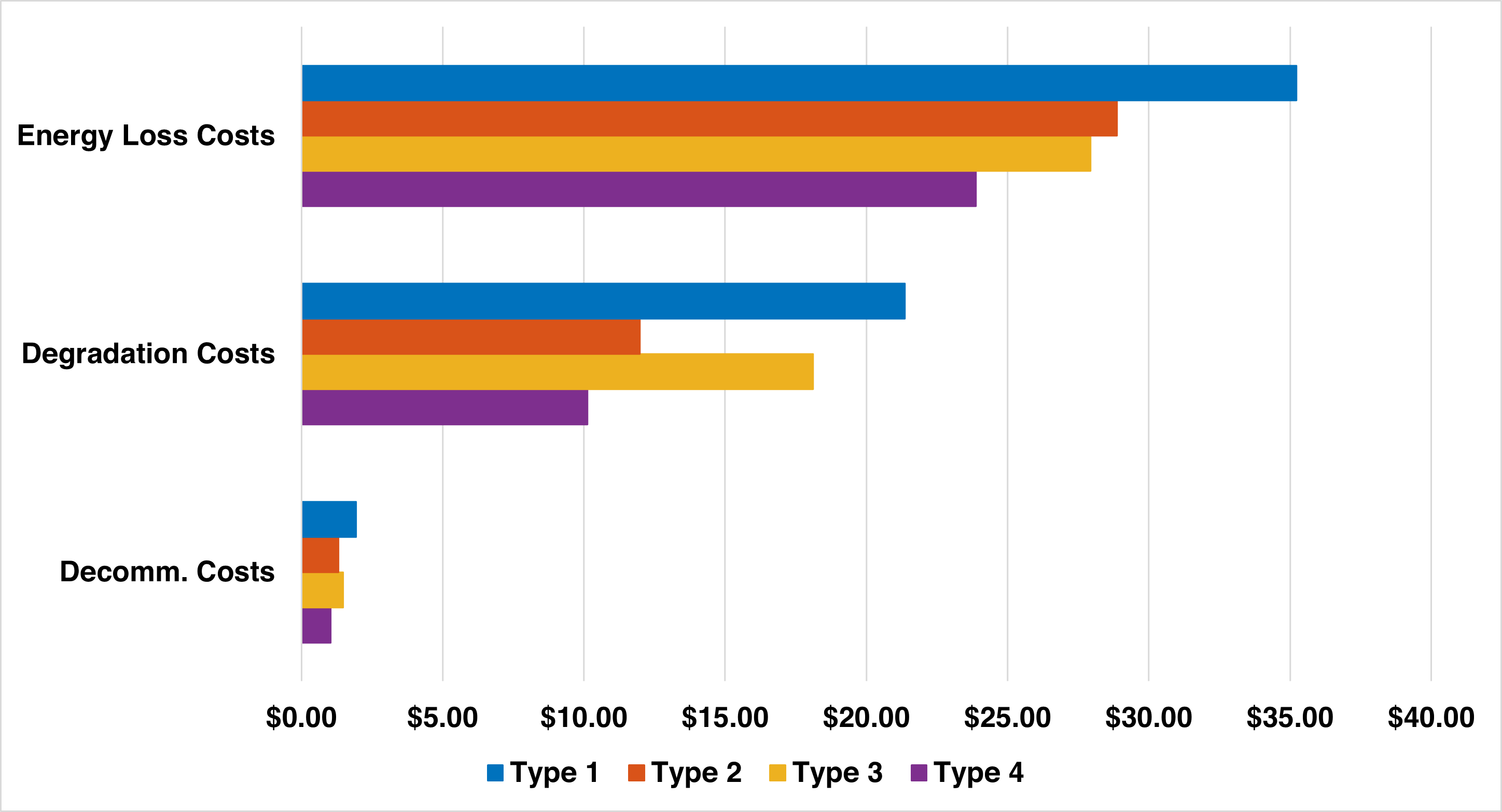} 
\caption{The cost analysis of the proposed approach for the first cycle.}
\label{Case1_2}
\end{figure}

Fig.~\ref{Case1_2} presents the cost breakdown for the proposed approach across the four battery pack types. Types 1 and 2 have the most overall costs, since their larger capacities imply more throughputs in charging/discharging. Type 4 contributes the least cost as it is the weakest among all in capacity, efficiency, and SoH, and thus used the least. For Types 1 and 3, the  percentage of the degradation cost relative to the energy loss cost is less than that for Types 2 and 4. This is because Types 1 and 3 have higher SoH when the second life begins and higher efficiency $\eta_{\textrm{ch/disch}}$.


\begin{figure}[!t]
\centering
\includegraphics[trim={0cm 0cm 0cm 0cm}, clip, width=8cm]{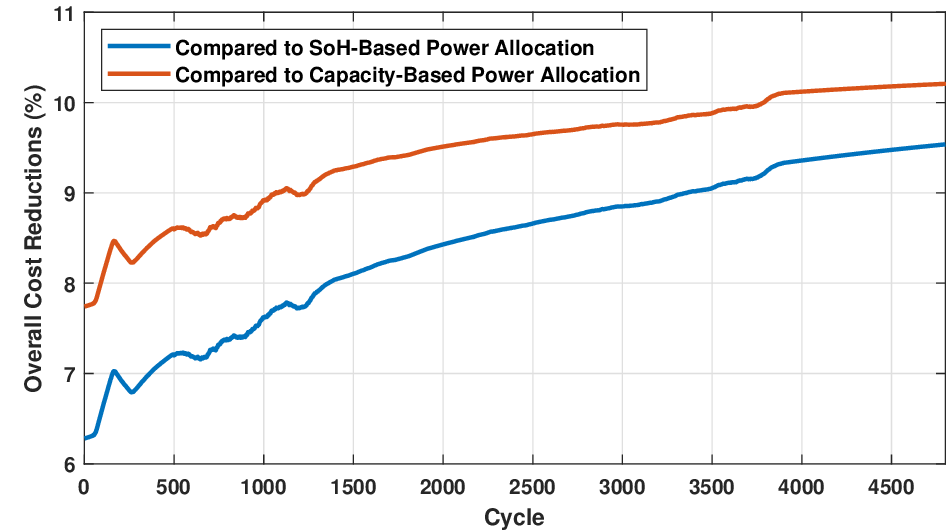}
\caption{The overall cost reductions of the proposed approach.}
\label{Case1Save}
\end{figure}

For a comprehensive assessment, we compute the overall cost reductions over 4800 cycles, as displayed in Fig.~\ref{Case1Save}. In the first cycle, our approach achieves approximately 6-7\% cost reduction compared to SoH-based and capacity-based approaches, as demonstrated above. This margin further increases to over 9-10\% in later cycles, underscoring the significance of economic optimal power management throughout the second life of the battery packs.

\begin{figure}[!t]
\centering
\includegraphics[trim={0cm 0cm 0cm 0cm}, clip, width=8cm]{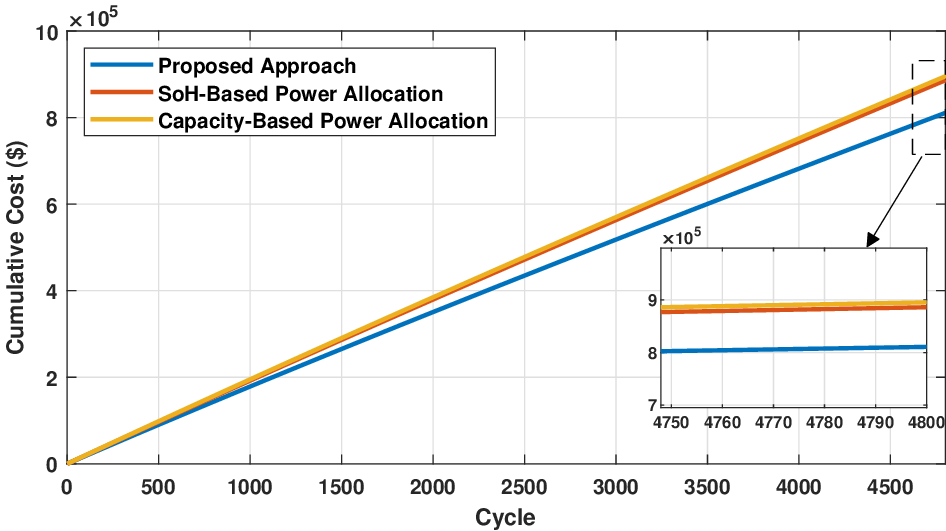}
\caption{The cumulative cost of operation under the three power allocation approaches.}
\label{Case1CumulativeCost}
\end{figure}

Fig.~\ref{Case1CumulativeCost} shows the cumulative costs over 4800 operation cycles for the three power allocation approaches. The proposed approach incurs \$810.8K, while the SoH-based and capacity-based approaches incur \$886.1K and \$895.3K, respectively, representing increases of about 9.2\% and 10.4\%.

\begin{figure}[!t]
\centering
\includegraphics[trim={1cm 0.5cm 1cm 0.2cm}, clip, width=\linewidth]{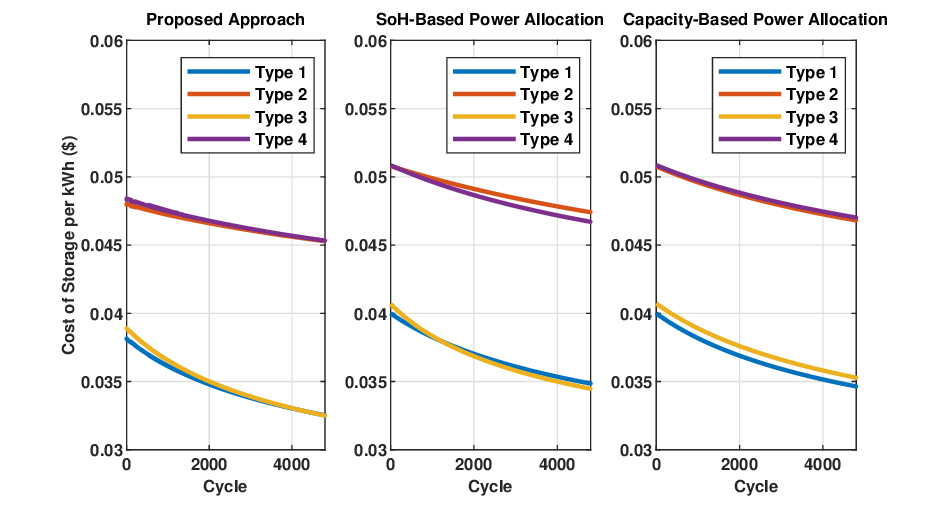}
\caption{The cost of storage under the three power allocation approaches.}
\label{Case1COE}
\end{figure}

Fig.~\ref{Case1COE} shows the average cost of storage (ACOS) per kWh for all battery pack types under the three power allocation approaches. In our proposed approach, Type-1 and Type-3 packs cost slightly below 0.04\$, whereas Type-2 and Type-4 packs are around 0.048\$. 
Additionally, the SoH-based and capacity-based approaches incur higher ACOS. The ACOS difference due to different power management approach emphasizes the necessity of optimal power management in second-life battery storage systems.

\subsection{Techno-Economic Analysis: Case 2}

In this case, we examine second-life battery packs comprising prismatic LiFePO4 cells investigated in \cite{AppliedEnergy-ZS-2014}. While maintaining a simulation setting consistent with Case 1, the aging model parameters in \eqref{AgingModel} are set according to \cite{AppliedEnergy-ZS-2014} as follows: $E=-15162$, $\beta=1516$, and $\zeta=0.824$. The pre-exponential factor $B_j$ is $0.1294$ for Type 1 and Type 3, and $0.1807$ for Type 2 and Type 4. We repeatedly apply the output power profile in Fig.~\ref{Pout} to the SL-BESS for 2400 cycles. In what follows, we report the key results of this case study.

\begin{figure}[!t]
\centering
\includegraphics[trim={1cm 0.5cm 1cm 0.2cm}, clip, width=\linewidth]{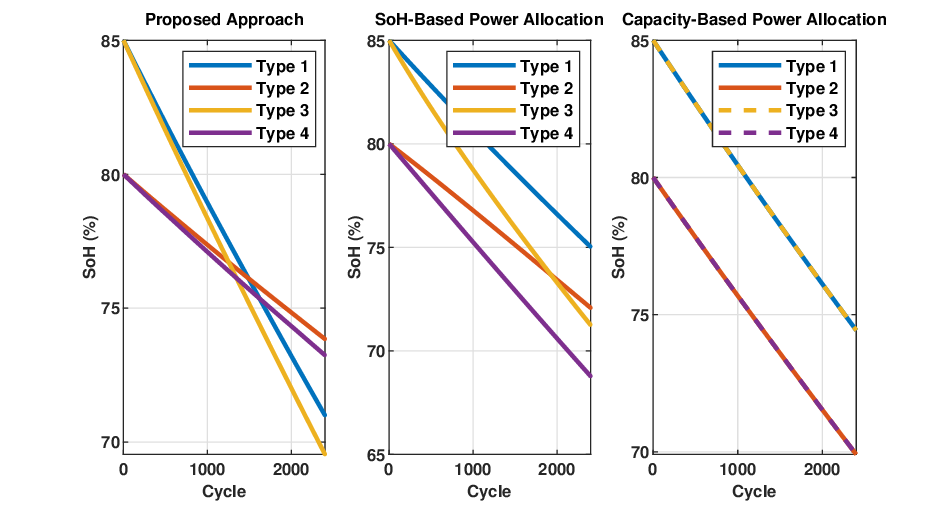}
\caption{The average degradation of different second-life battery pack types.}
\label{SoH2}
\end{figure}

\begin{figure}[!t]
\centering
\includegraphics[trim={0.25cm 0.25cm 0.25cm 0.25cm},clip,width=8.5cm]{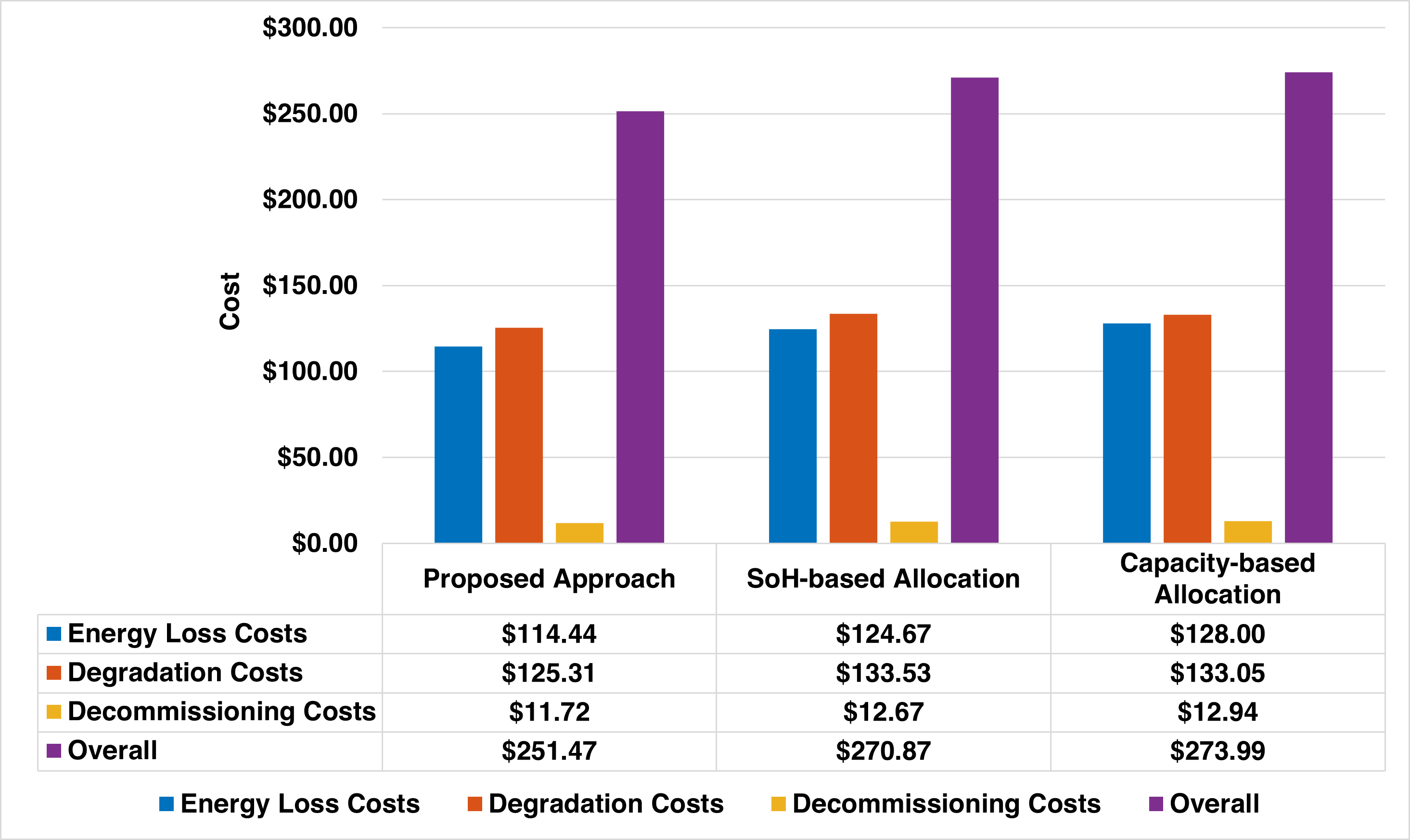} 
\caption{The cost analysis for different power allocation approaches.}
\label{Case2_1}
\end{figure}

\begin{figure}[!t]
\centering
\includegraphics[trim={1cm 0.5cm 1cm 0.2cm}, clip, width=\linewidth]{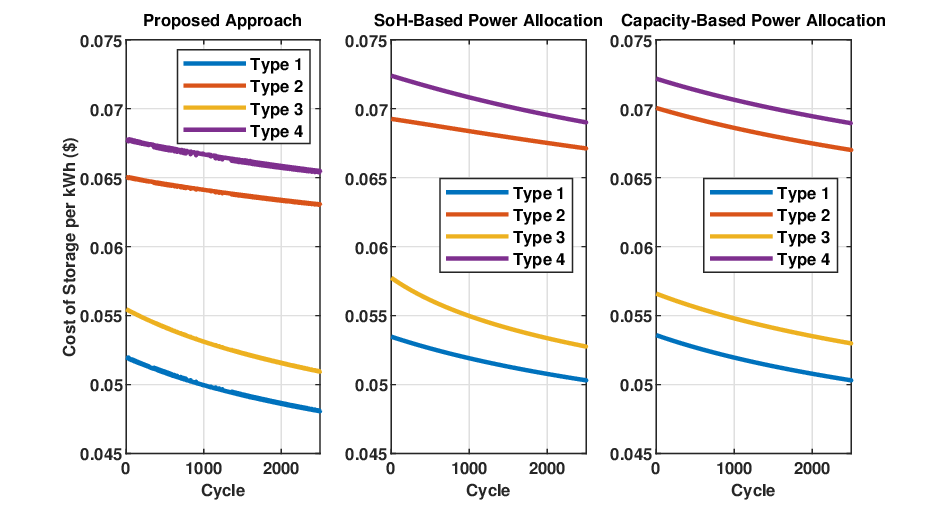}
\caption{The cost of storage under the three power allocation approaches.}
\label{Case2COE}
\end{figure}

Fig.~\ref{SoH2} displays the SoH evolution of the battery packs throughout the second life. Here, the degradation appears more linear and faster than that in Case 1, because of the new aging parameter setup for the considered prismatic cells. The battery packs now achieve almost the same degradation level in just 2400 cycles. The faster degradation has economic implications. The most obvious is that the total cost will become higher for the entire system. For example, Fig.~\ref{Case2_1} depicts the incurred costs at the 1200-th operating cycle for the three power allocation approaches, which all increase above \$250. As another implication, the degradation cost now takes a larger share in the overall cost structure. Fig.~\ref{Case2_1} shows that it overtakes the energy loss cost indeed, in a departure from what is shown in Case 1. Despite this, the proposed approach still outperforms the other two power allocation approaches economically by a significant margin. It yields the lowest cost of \$251.47, compared to \$270.84 and \$273.99 for the other approaches. The ACOS assessment in Fig.~\ref{Case2COE} further corroborates the analysis. The ACOS due to the proposed approach is around 0.06 \$/kWh, but well exceeds 0.065 \$/kWh for the other two.


\subsection{Discussion}
From the simulation study, we gain the following insights and perspectives.
\begin{itemize}
	\item {\em Economic power optimization is essential for SL-BESS.} Our simulation results consistently show that economic optimal power management improves the economic performance of SL-BESS by 6-10\% or even more. The non-trivial margin of improvement suggests that economic power optimization can play a significant role in driving down the costs of adopting SL-BESS. This notion, which has been overlooked in the literature, deserve being considered in future techno-economic analysis of SL-BESS projects and applications.
	\item {\em Selecting economical second-life battery packs.} The study also offers some guidelines for purchasing second-life battery packs towards maximum operating cost effectiveness. Based on our simulations, we introduce an index to assess the economic viability of a second-life battery pack as follows:
	\begin{equation}
		\textrm{EI} = \frac{C_{\textrm{cap}} + C_{\textrm{decom}}}{\eta\mathcal{Z}_{\textrm{SL}}}, \nonumber
	\end{equation}
	where $\mathcal{Z}_{\textrm{SL}}$ is the remaining total Ah throughput in the pack's second life. A lower $\textrm{EI}$ indicates greater battery pack economy. One is suggested to select second-life battery packs for their projects based on this index.
	\item {\em ACOS assessment.} The above simulations consider an operation time window of 12 hours. Relatively short as it is, we can use the results to make a rudimentary ACOS assessment for deploying the SL-BESS technology for grid energy storage. What is worth highlighting is that the ACOS in both Cases 1 and 2, though excluding factors like upfront investment, installation, and operational maintenance, is arguably close to the U.S. Department of Energy's target of achieving an ACOS of 0.05 \$/kWh \cite{DAYS}. We hence conclude that the SL-BESS technology is hopeful to become an economically viable means for grid energy storage, especially when optimal economic power management is applied.
	\item {\em Potential extensions of research.} Our study focuses on the problem of economic optimal power management for SL-BESS, and along this line of research, there are a diverse range of open problems. For instance, we can expand the work by developing more accurate degradation models that include calendar aging and depth-of-discharge. The sensitivity of the economic power optimization with respect to different parameters is worth exploring, though beyond the scope of this paper. Another interesting question is how to perform economic power management for SL-BESS in the contexts of demand response or economic dispatch to maximize grid-level economy.
\end{itemize}
\section{Conclusion}
The unfolding new era of EVs will lead to a surge in retired EV battery packs in the next decades, presenting a pressing need to repurpose them for second-life applications. We argue that economic power optimization is essential for SL-BESS consisting of heterogeneous second-life battery packs, with the goal of minimizing the economic costs while ensuring safe SL-BESS operation and matching the power supply with the demand. To achieve the goal, we propose an economic optimal power management approach, which is the first among its kind for SL-BESS. The proposed approach aims to minimize the economic costs while ensuring the safe SL-BESS operation and matching the power supply with the demand. Our approach considers three cost components: 1) energy loss costs due to the inefficiencies of the second-life battery packs, 2) degradation costs incurred by the loss of remaining lives of the second-life battery packs, and 3) decommissioning costs at the end of their lives. We accordingly present the cost models coupled with the dynamic models and then leverage the models to enable economic optimization. We conduct ample simulations under various use scenarios to validate the efficacy of the proposed approach. The simulation results show that the approach will enhance the economic performance of SL-BESS by 6-10\% or even more, depending on the specific operating conditions. 

\balance
\bibliographystyle{IEEEtran}
\scriptsize\bibliography{IEEEabrv,Bibliography/BIB}

\begin{thebibliography}{10}
\providecommand{\url}[1]{#1}
\csname url@samestyle\endcsname
\providecommand{\newblock}{\relax}
\providecommand{\bibinfo}[2]{#2}
\providecommand{\BIBentrySTDinterwordspacing}{\spaceskip=0pt\relax}
\providecommand{\BIBentryALTinterwordstretchfactor}{4}
\providecommand{\BIBentryALTinterwordspacing}{\spaceskip=\fontdimen2\font plus
\BIBentryALTinterwordstretchfactor\fontdimen3\font minus
  \fontdimen4\font\relax}
\providecommand{\BIBforeignlanguage}[2]{{%
\expandafter\ifx\csname l@#1\endcsname\relax
\typeout{** WARNING: IEEEtran.bst: No hyphenation pattern has been}%
\typeout{** loaded for the language `#1'. Using the pattern for}%
\typeout{** the default language instead.}%
\else
\language=\csname l@#1\endcsname
\fi
#2}}
\providecommand{\BIBdecl}{\relax}
\BIBdecl

\bibitem{McKinsey}
\BIBentryALTinterwordspacing
H.~Engel, P.~Hertzke, and G.~Siccardo. (2019, Apr.) Second-life {EV} batteries:
  The newest value pool in energy storage. [Online]. Available:
  \url{https://www.mckinsey.com/industries/automotive-and-assembly/our-insights/second-life-ev-batteries-the-newest-value-pool-in-energy-storage}
\BIBentrySTDinterwordspacing

\bibitem{Enviro-LCC-2019}
L.~C. Casals, B.~{Amante García}, and C.~Canal, ``Second life batteries
  lifespan: Rest of useful life and environmental analysis,'' \emph{Journal of
  Environmental Management}, vol. 232, pp. 354--363, 2019.

\bibitem{CellReports-JZ-2021}
J.~Zhu, I.~Mathews, D.~Ren, W.~Li, D.~Cogswell, B.~Xing, T.~Sedlatschek,
  S.~N.~R. Kantareddy, M.~Yi, T.~Gao, Y.~Xia, Q.~Zhou, T.~Wierzbicki, and M.~Z.
  Bazant, ``End-of-life or second-life options for retired electric vehicle
  batteries,'' \emph{Cell Reports Physical Science}, vol.~2, no.~8, p. 100537,
  2021.

\bibitem{Review-XG-2024}
X.~Gu, H.~Bai, X.~Cui, J.~Zhu, W.~Zhuang, Z.~Li, X.~Hu, and Z.~Song,
  ``Challenges and opportunities for second-life batteries: Key technologies
  and economy,'' \emph{Renewable and Sustainable Energy Reviews}, vol. 192, p.
  114191, 2024.

\bibitem{EnergyConversion-AB-2022}
A.~Bhatt, W.~Ongsakul, and N.~{Madhu M.}, ``Optimal techno-economic feasibility
  study of net-zero carbon emission microgrid integrating second-life battery
  energy storage system,'' \emph{Energy Conversion and Management}, vol. 266,
  p. 115825, 2022.

\bibitem{EVC-KA-2012}
A.~Keeli and R.~K. Sharma, ``Optimal use of second life battery for peak load
  management and improving the life of the battery,'' in \emph{IEEE
  International Electric Vehicle Conference}, 2012, pp. 1--6.

\bibitem{ECCE-KC-2015}
C.~Koch-Ciobotaru, A.~Saez-de Ibarra, E.~Martinez-Laserna, D.-I. Stroe,
  M.~Swierczynski, and P.~Rodriguez, ``Second life battery energy storage
  system for enhancing renewable energy grid integration,'' in \emph{IEEE
  Energy Conversion Congress and Exposition}, 2015, pp. 78--84.

\bibitem{ICIT-SI-2015}
A.~Saez-de Ibarra, E.~Martinez-Laserna, C.~Koch-Ciobotaru, P.~Rodriguez, D.-I.
  Stroe, and M.~Swierczynski, ``Second life battery energy storage system for
  residential demand response service,'' in \emph{IEEE International Conference
  on Industrial Technology}, 2015, pp. 2941--2948.

\bibitem{EEEIC-SA-2021}
A.~Soto, A.~Berrueta, P.~Zorrilla, A.~Iribarren, D.~H. Castillo, W.~E.
  Rodríguez, A.~J. Rodríguez, D.~T. Vargas, I.~R. Matias, P.~Sanchis, and
  A.~Ursúa, ``Integration of second-life battery packs for self-consumption
  applications: analysis of a real experience,'' in \emph{IEEE International
  Conference on Environment and Electrical Engineering and IEEE Industrial and
  Commercial Power Systems Europe}, 2021, pp. 1--6.

\bibitem{AppliedEnergy-MG-2021}
M.~Guo, Y.~Mu, H.~Jia, Y.~Deng, X.~Xu, and X.~Yu, ``Electric/thermal hybrid
  energy storage planning for park-level integrated energy systems with
  second-life battery utilization,'' \emph{Advances in Applied Energy}, vol.~4,
  p. 100064, 2021.

\bibitem{IET-DY-2022}
Y.~Deng, Y.~Zhang, F.~Luo, and Y.~Mu, ``Hierarchical energy management for
  community microgrids with integration of second-life battery energy storage
  systems and photovoltaic solar energy,'' \emph{IET Energy Systems
  Integration}, vol.~4, no.~2, pp. 206--219, 2022.

\bibitem{TSE-DY-2021}
------, ``Operational planning of centralized charging stations utilizing
  second-life battery energy storage systems,'' \emph{IEEE Transactions on
  Sustainable Energy}, vol.~12, no.~1, pp. 387--399, 2021.

\bibitem{AppliedEnergy-ZS-2019}
Z.~Song, S.~Feng, L.~Zhang, Z.~Hu, X.~Hu, and R.~Yao, ``Economy analysis of
  second-life battery in wind power systems considering battery degradation in
  dynamic processes: Real case scenarios,'' \emph{Applied Energy}, vol. 251, p.
  113411, 2019.

\bibitem{SelectedTPEL-RM-2023}
M.~Rasheed, R.~Hassan, M.~Kamel, H.~Wang, R.~Zane, S.~Tong, and K.~Smith,
  ``Active reconditioning of retired lithium-ion battery packs from electric
  vehicles for second life applications,'' \emph{IEEE Journal of Emerging and
  Selected Topics in Power Electronics}, pp. 1--1, 2023, {I}n Press.

\bibitem{Energies-AM-2016}
M.~Abdel-Monem, O.~Hegazy, N.~Omar, K.~Trad, S.~De~Breucker, P.~Van
  Den~Bossche, and J.~Van~Mierlo, ``Design and analysis of generic energy
  management strategy for controlling second-life battery systems in stationary
  applications,'' \emph{Energies}, vol.~9, no.~11, 2016.

\bibitem{TPEL-LC-2020}
C.~Liu, N.~Gao, X.~Cai, and R.~Li, ``Differentiation power control of modules
  in second-life battery energy storage system based on cascaded {H}-bridge
  converter,'' \emph{IEEE Transactions on Power Electronics}, vol.~35, no.~6,
  pp. 6609--6624, 2020.

\bibitem{TIE-MN-2015}
N.~Mukherjee and D.~Strickland, ``Control of second-life hybrid battery energy
  storage system based on modular boost-multilevel buck converter,'' \emph{IEEE
  Transactions on Industrial Electronics}, vol.~62, no.~2, pp. 1034--1046,
  2015.

\bibitem{TSG-ZQ-2024}
Q.~Zhao, K.~Liao, J.~Yang, Z.~He, and Y.~Xu, ``Aging rate equalization strategy
  for battery energy storage systems in microgrids,'' \emph{IEEE Transactions
  on Smart Grid}, vol.~15, no.~1, pp. 136--148, 2024.

\bibitem{APEC-LC-2020}
C.~Lamoureux, Z.~Gong, M.~Nasr, S.~A. Assadi, K.~Gupta, D.~Galatro, O.~Tayyara,
  C.~da~Silva, C.~Amon, and O.~Trescases, ``Electrochemical impedance
  spectroscopy based power-mix control strategy for improved lifetime
  performance in second-life battery systems,'' in \emph{IEEE Applied Power
  Electronics Conference and Exposition}, 2020, pp. 3444--3451.

\bibitem{TTE-FA-2023}
A.~Farakhor, D.~Wu, Y.~Wang, and H.~Fang, ``A novel modular, reconfigurable
  battery energy storage system: Design, control, and experimentation,''
  \emph{IEEE Transactions on Transportation Electrification}, vol.~9, no.~2,
  pp. 2878--2890, 2023.

\bibitem{2021-IECON-FA}
A.~Farakhor and H.~Fang, ``A novel modular, reconfigurable battery energy
  storage system design,'' in \emph{47th Annual Conference of the IEEE
  Industrial Electronics Society}, 2021, pp. 1--6.

\bibitem{NetworkSystems-YT-2019}
T.~Yang, D.~Wu, H.~Fang, W.~Ren, H.~Wang, Y.~Hong, and K.~H. Johansson,
  ``Distributed energy resource coordination over time-varying directed
  communication networks,'' \emph{IEEE Transactions on Control of Network
  Systems}, vol.~6, no.~3, pp. 1124--1134, 2019.

\bibitem{SAE-NJ-2015}
J.~S. Neubauer, E.~Wood, and A.~Pesaran, ``Second life for electric vehicle
  batteries: Answering questions on battery degradation and value,'' \emph{SAE
  International Journal of Materials and Manufacturing}, vol.~8, no.~2, pp.
  544--553, 2015.

\bibitem{TTE-VW-2022}
W.~Vermeer, G.~R. Chandra~Mouli, and P.~Bauer, ``A comprehensive review on the
  characteristics and modeling of lithium-ion battery aging,'' \emph{IEEE
  Transactions on Transportation Electrification}, vol.~8, no.~2, pp.
  2205--2232, 2022.

\bibitem{TSG-LS-2024}
S.~Li, P.~Zhao, C.~Gu, Y.~Xiang, S.~Bu, E.~Chung, Z.~Tian, J.~Li, and S.~Cheng,
  ``Factoring electrochemical and full-lifecycle aging modes of battery
  participating in energy and transportation systems,'' \emph{IEEE Transactions
  on Smart Grid}, pp. 1--1, 2024.

\bibitem{PowerSources-JZ-2021}
J.~Zhu, M.~Knapp, D.~R. Sørensen, M.~Heere, M.~S. Darma, M.~Müller,
  L.~Mereacre, H.~Dai, A.~Senyshyn, X.~Wei, and H.~Ehrenberg, ``Investigation
  of capacity fade for 18650-type lithium-ion batteries cycled in different
  state of charge ({S}o{C}) ranges,'' \emph{Journal of Power Sources}, vol.
  489, p. 229422, 2021.

\bibitem{SmartGrid-LM-2017}
M.~Liu, W.~Li, C.~Wang, M.~P. Polis, L.~Y. Wang, and J.~Li, ``Reliability
  evaluation of large scale battery energy storage systems,'' \emph{IEEE
  Transactions on Smart Grid}, vol.~8, no.~6, pp. 2733--2743, 2017.

\bibitem{InternationalPowerElectronics-OS-2012}
S.~Onori, P.~Spagnol, V.~Marano, Y.~Guezennec, and G.~Rizzoni, ``A new life
  estimation method for lithium-ion batteries in plug-in hybrid electric
  vehicles applications,'' \emph{International Journal of Power Electronics},
  vol.~4, no.~3, pp. 302--319, 2012.

\bibitem{PowerSources-JW-2011}
J.~Wang, P.~Liu, J.~Hicks-Garner, E.~Sherman, S.~Soukiazian, M.~Verbrugge,
  H.~Tataria, J.~Musser, and P.~Finamore, ``Cycle-life model for
  graphite-{L}i{F}e{PO}4 cells,'' \emph{Journal of Power Sources}, vol. 196,
  no.~8, pp. 3942--3948, 2011.

\bibitem{2016-TSTE-PC}
C.~Pinto, J.~V. Barreras, E.~Schaltz, and R.~E. Ara\'{u}jo, ``Evaluation of
  advanced control for \uppercase{L}i-ion battery balancing systems using
  convex optimization,'' \emph{IEEE Transactions on Sustainable Energy},
  vol.~7, no.~4, pp. 1703--1717, 2016.

\bibitem{PNNL-WV-2022}
V.~Viswanathan, K.~Mongird, R.~Franks, and R.~Baxter, ``2022 grid energy
  storage technology cost and performance assessment,'' Pacific Northwest
  National Laboratory, Tech. Rep. PNNL-33283, Aug. 2022.

\bibitem{AppliedEnergy-ZS-2014}
Z.~Song, J.~Li, X.~Han, L.~Xu, L.~Lu, M.~Ouyang, and H.~Hofmann,
  ``Multi-objective optimization of a semi-active battery/supercapacitor energy
  storage system for electric vehicles,'' \emph{Applied Energy}, vol. 135, pp.
  212--224, 2014.

\bibitem{DAYS}
{U.S. Department of Energy}, ``Duration addition to electricity storage
  ({DAYS}) overview,''
  \url{https://arpa-e.energy.gov/sites/default/files/documents/files/DAYS_ProgramOverview_FINAL.pdf},
  Advanced Research Projects Agency-Energy (ARPA-E), Tech. Rep.

\end{thebibliography}

\vfill

\end{document}